\documentclass[graphics, twocolumn, usenatbib]{mn2e}
\usepackage{natbib} 
\usepackage{epsfig} 
\usepackage{graphicx} 
\usepackage{color}
\usepackage{aas_macros} 
\usepackage{amssymb}
\usepackage{amsmath}
\usepackage[title]{appendix}

\newcommand{\lya}{Lyman-$\alpha$~}

\newcommand{\enzo}{\texttt{Enzo~}}
\newcommand{\grackle}{\texttt{Grackle-2.1~}}
\newcommand{\gracklec}{\texttt{Grackle-2.1}}

\newcommand{\mpch} {\rm $h^{-1}$ Mpc\,\,} 
\newcommand{\kpch} {\rm $h^{-1}$ kpc\,\,} 
\newcommand{\msolar} {$\rm{M_{\odot}}~$}
\newcommand{\msolarc} {$\rm{M_{\odot}}$}
\newcommand{\zsolar} {$\rm{Z_{\odot}}~$}

\newcommand{\molH} {$\rm{H_2}$~}
\newcommand{\molHc} {$\rm{H_2}$}
\newcommand{\J} {$\rm{10^{-21}\ erg\ cm^{-2}\ s^{-1}\ Hz^{-1}\ sr^{-1}}$}

\newcommand{\healpix} {\texttt{HEALPix~}}

\def\etal{{\it et al.}~}
\begin{document}
\title[]{Forming Super-Massive Black Hole Seeds under the Influence of a Nearby 
 Anisotropic Multi-Frequency Source}

\author[J.A. Regan \etal] 
{John A. Regan$^{1, 2}$\thanks{E-mail:john.a.regan@durham.ac.uk}, Peter H. Johansson$^{2}$ 
\& John H. Wise$^{3}$ \\ \\
$^1$Institute for Computational Cosmology, Durham University, South Road, Durham, UK, DH1 3LE \\
$^2$Department of Physics, University of Helsinki, Gustaf H\"allstr\"omin katu 2a,
FI-00014 Helsinki, Finland \\
$^3$Center for Relativistic Astrophysics, Georgia Institute of Technology, 837 State Street, 
Atlanta, GA 30332, USA
\\}


\maketitle

\begin{abstract} 
The photo-dissociation of \molH by a nearby anisotropic source of radiation 
is seen as a critical component in creating an environment in which a direct collapse
black hole may form. Employing radiative transfer we model the effect of multi-frequency 
(0.76 eV - 60 eV) radiation on a collapsing halo at high redshift. We vary both the shape of 
the spectrum which emits the radiation and the distance to the emitting galaxy. We use blackbody 
spectra with temperatures of $\rm{T = 10^4\ K}$ and $\rm{T = 10^5\ K}$ and a realistic stellar 
spectrum.  We find that an optimal zone exists between 1 kpc and 4 kpc from the emitting galaxy. 
If the halo resides 
too close to the emitting galaxy the photo-ionising radiation creates a large HII region 
which effectively disrupts the collapsing halo, too far from the source and the radiation 
flux drops below the level of the expected background and the \molH fraction remains too high. 
When the emitting galaxy is initially placed between 1 kpc and 2 kpc from the collapsing halo, 
with a spectral shape consistent with a star-forming high redshift galaxy, then a large central 
core forms. The mass of the central core is between 5000 and 10000 \msolar at a temperature of 
approximately 1000 K. This core is however surrounded by a reservoir of hotter gas at approximately 8000 K
which leads to mass inflow rates of the order of $\sim 0.1$ \msolar yr$^{-1}$.

\end{abstract}



\section{Introduction} \label{Sec:Introduction}
The discovery of a significant number of quasars ($\sim 40$) at redshifts greater than six hosting 
black holes with masses exceeding $\gtrsim 10^9$ \msolar \citep{Fan_2006b, Mortlock_2011, 
Venemans_2013, Wu_2015} has challenged our understanding of how super-massive black holes (SMBHs)
can form. The most straightforward mechanism is to assume that SMBHs grow through accretion and 
possibly mergers of remnant black holes from Population III (Pop III) collapse at the 
end of their rather short lifetimes. However, this argument suffers from numerous obstacles, with
the prime issue being a timescale argument. The growth by spherical accretion of a black hole is governed 
by the Eddington limit 
\begin{equation}
M(t) = M(t_0) \ \rm{exp \Big({1 - \epsilon \over \epsilon } {t \over t_{edd}}\Big)}
\end{equation}
where $\epsilon$ is the radiative efficiency, M(t) is the mass at time t and 
$t_{\rm{edd}}$ is the Eddington time,  $t_{\rm{edd}}$ is approximately 450 Myrs taking $\epsilon = 0.1$. 
Therefore a stellar mass black hole forming at 500 Myrs has of the order of 10 or so e-folding times to 
reach a mass of $10^9$ \msolarc, where the growth is due to radiatively efficient accretion onto 
the black hole. This, coupled with the fact that the initial mass function (IMF) of the first stars 
is hotly debated \citep{Stacy_2010, Greif_2011, Clark_2011, Bromm_2013, Hirano_2014} and that subsequent 
accretion onto the first stellar mass black holes is likely to be inefficient 
\citep{Johnson_2007, Milosavljevic_2009, Alvarez_2009, Hosokawa_2011} makes this mechanism for 
forming early SMBHs unattractive and difficult to reconcile with current observations of 
super-massive black holes at high redshift.\\
\indent The direct collapse mechanism circumvents the above limitations to some extent by 
forming large seed black holes making growth to super-massive size by a redshift of six
achievable. The pathway to producing a large seed is currently unclear with numerous
avenues under investigation \citep{Bromm_2003, Wise_2008a, Regan_2009b, Regan_2009, Tseliakhovich_2010,
Inayoshi_2012, Agarwal_2013, Latif_2013c, Tanaka_2014, Agarwal_2014b, Mayer_2014, Regan_2014a, Inayoshi_2015}. 
Regardless of the final outcome, the direct collapse mechanism requires that the gas cloud that 
eventually collapses to form a massive black hole seed is hotter than the gas cloud that produces 
the first stars. The increased temperature elevates the Jeans mass thus allowing a larger object 
to initially form. In order to keep the temperature of the gas high, cooling must be somehow 
disrupted. Assuming the gas to be metal-free this means that the availability of \molH must be 
reduced. This can be achieved either through photo-dissociation or collisional dissociation. \\
\indent Collisional dissociation of \molH (\molH + H $\rightarrow$ 3 H) is effective for gas 
of a primordial composition and high temperature satisfying the criteria of the ``zone of no-return'' 
\citep{Visbal_2014}. \cite{Inayoshi_2012} suggested that cold accretion shocks may provide a 
pathway to collisionally dissociate \molH during gravitational collapse, however, 
\cite{Fernandez_2014} demonstrated, through numerical simulations, that in the absence of a 
photo-dissociating background this method is difficult to achieve in practice as the 
collisional processes tend to operate at the virial radius and not in the centre of the halo. \\
\indent Photo-dissociation of \molH has been studied by several authors as a viable means 
of disrupting \molH cooling at high redshift where metal cooling is unavailable 
\citep{Omukai_2001, Oh_2002, Bromm_2003, Shang_2010, Latif_2014a, Latif_2014b, Latif_2015}. 
In this case radiation in the Lyman-Werner (LW) band with energies between 11.2 and 13.6 eV 
is able to dissociate \molH via the two step Solomon process \citep{Field_1966, Stecher_1967}. 
The process operates by exciting the molecule from the electronic ground state, 
$\rm{X^1 \Sigma_g^{+}}$, to the $\rm{B^1 \Sigma_u^{+}}$ or $\rm{C^1 \Gamma_u}$ state. These are the 
Lyman and Werner states of \molHc. The subsequent decay to the ground state then leads to the 
dissociation of the molecule in 15\% of cases. The Solomon process can therefore be written as
\begin{align}
\mathrm{H_2 + \gamma  \rightarrow H_2^{*}} \\
\mathrm{H_2^{*}      \rightarrow H + H + \gamma}
\end{align}
Lower energy radiation can also influence the \molH abundances by photo-detaching the 
intermediary ion H$^-$. The primary route to \molH formation is through the reaction
\begin{equation}
\mathrm{H + H^- \rightarrow H_2 + e^-}
\end{equation}
Therefore by photo-detaching the electron from the  H$^-$ ion the formation rate of \molH is 
severely compromised. 
\begin{equation}
\mathrm{H^- + \gamma \rightarrow H + e^-}
\end{equation}
where the photo-detachment threshold is approximately 0.76 eV. Radiation in the infrared band is 
therefore very effective at destroying the intermediary required for \molH formation. Finally, 
H$_2^{+}$ is also an intermediary for forming \molHc. 
\begin{equation}
\mathrm{H_2^{+} + H     \rightarrow H_2^{*} + H^+}
\end{equation}
H$_2^{+}$ is destroyed by radiation between 
approximately 0.1 eV and 25 eV \citep{Stancil_1994}. 
\begin{equation}
\mathrm{H_2^{+} + \gamma \rightarrow H + H^+}
\end{equation}

\begin{table*}
\centering
\caption{Reaction Network}
\begin{tabular}{ | l | l | l | l | }
\hline 
\textbf{\em $\rm{No.}$}
& \textbf{\em $\rm{Reaction}$} & \textbf{\em $\rm{Notes}$}
& \textbf{\em $\rm{Reference}$} \\

\hline 

1. &  $\rm{H + e             \rightarrow   H^+ + 2e}$        &  - &  1, 2\\
2. &  $\rm{H^+ + e           \rightarrow   H + \gamma}$      &  Case B Recombination &  1, 3\\
3. &  $\rm{He + e           \rightarrow   He^+ + \gamma}$   &  - &  1, 2 \\
4. &  $\rm{He^+ + e          \rightarrow  He + \gamma}$     & Effective He$^+$ Recombination Rate$^*$ & 1, 4 \\
5. &  $\rm{He^+ + e          \rightarrow  He^{++} + 2e}$     & -  & 5 \\
6. &  $\rm{He^{++} + e        \rightarrow  He^+ + \gamma}$   &  Case B Recombination &  5 \\
7. &  $\rm{H + e             \rightarrow   H^- + \gamma}$    &  - &  1, 6\\
8. &  $\rm{H^- + H           \rightarrow   H_{2} + e}$        &  - &  7, 8\\
9. &  $\rm{H + H^+           \rightarrow   H_{2}^+ + \gamma}$ &  - &  7, 9, 10\\
10. &  $\rm{H_{2}^+ + H        \rightarrow   H_2 + H^+}$       &  - &  1, 11\\
11. &  $\rm{H_2 + H^+         \rightarrow   H_{2}^{+} + H}$     &    - &  1, 12\\
12. &  $\rm{H_2 + H           \rightarrow   3H}$              &  - &  7, 13\\
13. &  $\rm{H^- + e           \rightarrow   H + 2e}$          &  - &  1, 2\\
14. &  $\rm{H^- + H          \rightarrow   2H + e}$          &  - &  1, 2\\
15. &  $\rm{H^- + H^+        \rightarrow   2H }$             &  - &  1, 14\\
16. &  $\rm{H^- + H^+        \rightarrow   H_{2}^{+} + e}$     &  - &  1, 15\\
17. &  $\rm{H_{2}^{+} + e     \rightarrow   2H}$              &  - &  5, 7, 10, 16 \\
18. &  $\rm{H + H + H       \rightarrow   H + H_2}$         &  - &  7,  17\\
19. &  $\rm{H + H           \rightarrow   H + H^+ + e}$     &  - &  7,  18, 19\\
20. &  $\rm{H + He          \rightarrow   H^+ + He + e}$    &  - &  7,  20\\
21. &  $\rm{H_{2}^{+} + H    \rightarrow   H^+ + H + H}$     &  - &  7,  10, 20\\
22. &  $\rm{H_{2}^{+} + He   \rightarrow   HeH^+ + H}$       &  - &  1,  21\\
23. &  $\rm{H_{2} + He      \rightarrow   H + H + He}$      &  - &  1,  22, 23\\
24. &  $\rm{HeH^+ + H       \rightarrow   H_{2}^+ + He}$    &  - &  1,  24\\
25. & $\rm{H^- + He         \rightarrow   H + He + e}$     &  - &  1\\
26. & $\rm{He + H^+         \rightarrow   HeH^+ + \gamma}$ &  - &  1, 25\\
27. & $\rm{HeH^+ + e        \rightarrow   He + H}$         &  - &  1, 26\\
28. & $\rm{H_2 + \gamma     \rightarrow   H + H}$          & Use fitting function from \cite{Wolcott-Green_2011} & \\
29. & $\rm{H^- + \gamma     \rightarrow   H + e}$          & Photo Detachment & 27\\
30. & $\rm{H_2^{+} + \gamma  \rightarrow   H + H^+}$        & Photo-Dissociation & 28 \\ 
31. &  $\rm{H + \gamma      \rightarrow   H^+ + e}$        & Hydrogen Ionisation & \\ 
32. &  $\rm{He + \gamma     \rightarrow   He^+ + e}$       & Helium Ionisation & \\ 
33. &  $\rm{He^+ + \gamma   \rightarrow   He^{++} + e}$     & Double Helium Ionisation & \\ 

\hline

\end{tabular}
\parbox[t]{0.9\textwidth}{\textit{Notes:} The 33 species reaction network used in our modified
version of \grackle. References: 1. \cite{GloverSavin_2009} 2. \cite{Janev_1987} 
3. \cite{Ferland_1992} 4. \cite{Hummer_1998}  5. \cite{Abel_1997} 6. \cite{Wishart_1979} 
7. \cite{Glover_2015a} 8. \cite{Kreckel_2010} 9. \cite{Latif_2015} 10. \cite{Coppola_2011} 
11. \cite{Karpas_1979} 12. \cite{Savin_2004} 13. \cite{Martin_1996} 14. \cite{Croft_1999}. 
15. \cite{Poulaert_1978} 16. \cite{Scheider_1994} 17.\cite{Forrey_2013}
18. \cite{Lenzuni_1991} 19. \cite{Omukai_2000} 20. \cite{Krstic_2003} 21. \cite{Black_1981}
22. \cite{Dove_1987} 23. \cite{Walkauskas_1975} 24. \cite{Linder_1995} 25. \cite{Jurek_1995}
26. \cite{Guberman_1994} 27. \cite{Tegmark_1997} 28. \cite{Stancil_1994}. 
$^*$This is a linear combination of Case A, Case B and Dielectric contributions as described in \cite{GloverSavin_2009}.
}

\label{Table:reaction_network}

\end{table*}
\indent Numerous studies have been undertaken to uncover the flux required to disrupt \molH formation
to the extent that a large central object can form within a halo cooled predominantly by 
atomic hydrogen \citep[e.g.][]{Omukai_2001, Shang_2010, Wolcott-Green_2011} with the general 
consensus being that an intensity of approximately 1000 J$_{21}$\footnote{J$_{21}$ is defined as \J}
is required for radiation with a blackbody spectrum of $10^5$ K, with an intensity of closer to 100 
J$_{21}$ required for radiation with a blackbody spectrum of $10^4$ K. However, more recent studies
have called into question the appropriateness of assuming a blackbody spectrum 
\citep{Sugimura_2014, Agarwal_2015a}, when instead, a more realistic spectral energy distribution 
(SED) is what is required. \cite{Agarwal_2015b} has also noted 
that trying to determine a single value of J$\rm{_{crit}}$ is likely to be very difficult
given the dependence of  J$\rm{_{crit}}$ on the distance to the nearby radiation source(s), 
and its spectral shape and evolution. Furthermore, as discussed by \cite{Latif_2015b} an 
isothermal collapse is not necessarily required to form a super-massive star and subsequently 
a direct collapse black hole (DCBH). In this case then finding a single value of  J$\rm{_{crit}}$ 
becomes even more challenging. \\
\indent Rather the focus should centre on modelling the direct collapse under realistic cosmological 
conditions. Using the results from high resolution simulations of the early universe and using them 
to determine a realistic SED turns earlier approaches on their heads. Instead of trying to determine 
a value for the intensity, $J$, we should model the effect realistic sources can have 
and study the viability of the direct collapse model under realistic cosmological 
conditions as found in the very early universe.\\
\indent In this work we focus on the key component of an anisotropic source. Building on the work of 
\cite{Regan_2014b} (hereafter R14) we evaluate the collapse of a high redshift gas cloud under 
the influence of a nearby anisotropic source. We have included radiation from 0.76 
eV up to 60 eV allowing us to probe the impact from a much more realistic radiation source. 
This is in comparison to R14, where only the effects of radiation in the LW band were included.
Furthermore, we have updated our chemical model based on the work of \cite{Glover_2015a}. We model 
the irradiating source as a blackbody with effective temperatures of $\rm{T_{eff} = 10^4 K}$ and 
$\rm{T_{eff} = 10^5 K}$ and also using a realistic spectral energy distribution generated
using the stellar population synthesis models of \cite{Bruzual_2003}. The parameters for creating 
the SED is based on the star formation rates and stellar masses found in the 
``Renaissance Simulations'' of \cite{Chen_2014}. We have imposed a cutoff at energies greater than 60 eV 
in this study thus ignoring the effects of X-rays in this case. The effects (both positive and negative) 
of X-ray radiation have been examined by \cite{Inayoshi_2011, Inayoshi_2015b} and \cite{Latif_2015}. We will also
examine this important component in an upcoming study \citep{Regan_2016b} but this study focuses
solely on the effects of the stellar component. \\
\indent The paper is laid out as follows: in \S \ref{Sec:NumericalSetup} we describe the 
numerical approach used including the halo setup, the chemical model and radiation 
prescription employed; in \S \ref{Sec:Results} we describe the results of our 
numerical simulations; in \S \ref{Sec:Discussion} we discuss the importance of the results 
and in \S \ref{Sec:Conclusions} we present our conclusions.  
Throughout this paper we  assume a standard $\Lambda$CDM cosmology with the following parameters 
\cite[based on the latest Planck data]{Planck_2014}, $\Omega_{\Lambda,0}$  = 0.6817, 
$\Omega_{\rm m,0}$ = 0.3183, $\Omega_{\rm b,0}$ = 0.0463, $\sigma_8$ = 0.8347 and $h$ = 0.6704. 
We further assume a spectral index for the primordial density fluctuations of $n=0.9616$.\\

\begin{table*}
\centering
\caption{Radiation Source}
\begin{tabular}{ | l | c | c | l | c | c | c | c |}
\hline 
\textbf{\em {Sim Name}} &
\textbf{\em {Init. Dist. (kpc)}} & \textbf{\em Spectrum} & \textbf{\em{z$_{\rm{coll}}$}} 
& \textbf{\em{Final Dist. (kpc)}} & \textbf{\em{T$_{\rm{vir}}$ (K)}} & \textbf{\em{M$_{200}$ (\msolarc)}} 
& \textbf{\em{M$_{\rm{core}}$ (\msolarc)}} \\
\hline 
Ctrl   & -   & -                 & z = 32.18  & -   & 1784 & $1.12 \times 10^6$ & 2010  \\
05-T4  & 0.5 & BB (T = $10^4$ K) & z = 30.08 & 0.8 & 3129 & $ 2.90 \times 10^6$ & 5625   \\
1-T4   & 1.0 & BB (T = $10^4$ K) & z = 30.86 & 1.4 & 2282 & $ 1.73 \times 10^6$ & 3686  \\
2-T4   & 2.0 & BB (T = $10^4$ K) & z = 31.63 & 2.6 & 2003 & $ 1.37 \times 10^6$ & 2633  \\
05-T5  & 0.5 & BB (T = $10^5$ K) & z = 21.50 & 5.0 & 9152 & $ 2.34 \times 10^7$ & 3494  \\
1-T5   & 1.0 & BB (T = $10^5$ K) & z = 22.81 & 5.3 & 8003 & $ 1.75 \times 10^7$ & 3537  \\
2-T5   & 2.0 & BB (T = $10^5$ K) & z = 29.17 & 2.8 & 3844 & $ 4.10 \times 10^6$ & 8132  \\
05-SSED & 0.5 & Stellar SED       & z = 21.70 & 4.9 & 10721 & $2.92 \times 10^7$ & 6146 \\
1-SSED  & 1.0 & Stellar SED       & z = 25.25 & 1.9 & 6224 & $1.04 \times 10^7$ & 9476  \\
2-SSED  & 2.0 & Stellar SED       & z = 28.67 & 2.9 & 4225 & $4.84 \times 10^6$ & 7269  \\
4-SSED  & 4.0 & Stellar SED       & z = 29.97 & 5.4 & 3181 & $2.96 \times 10^6$ & 6117  \\
8-SSED  & 8.0 & Stellar SED       & z = 30.87 & 10.4 & 2276 & $1.71 \times 10^6$ & 3889 \\
12-SSED & 12.0 & Stellar SED      & z = 31.44 & 15.2 & 2162 & $1.55 \times 10^6$ & 3243 \\
\hline

\end{tabular}
\parbox[t]{0.9\textwidth}{\textit{Notes:} Each model is run with the radiation source at an 
initial distance from the centre of the collapsing halo of 0.5, 1.0 and 2.0 kpc (physical). 
The initial distance is the distance at z = 40. 
For each of these models the spectrum is varied between a blackbody spectrum with an effective 
temperature of T = $10^4$ K (BB1e4) and T = $10^5$ K (BB1e5) and a stellar SED 
(maximum photon energy = 60 eV). Further simulations with the source placed at distances of 4.0, 
8.0 and 12.0 kpc are run for the stellar spectrum only. Finally, a control simulation (Ctrl) is 
run with no radiation field present. All distances are in physical kpc unless explicitly stated. 
The core mass in the final column denotes the baryonic mass inside a 1 pc radius around the 
densest point. 
}

\label{Table:radiation_particle}
\end{table*}


\section{Numerical Setup} \label{Sec:NumericalSetup}
\noindent We have  used the publicly available adaptive mesh refinement
(AMR) code \texttt{Enzo~}\citep{Enzo_2014}\footnote{http://enzo-project.org/}. In particular we use
version 3.0\footnote{Changeset: 7f49adb4c9b4} which is
the bleeding edge version of the code incorporating a range of new features. We created a fork
off the 3.0 mainline and included improved support for radiative transfer based on the Moray 
implementation of \cite{WiseAbel_2011} and chemical modelling using the \texttt{Grackle} library. \\
\indent All simulations are run within a box of 1 \mpch (comoving), the root grid size is $256^3$ and 
we employ three levels of nested grids. The grid nesting and initial conditions are created using 
MUSIC \citep{Hahn_2011}. Within the most refined region (i.e. level 3) the dark matter particle mass 
is $\sim$ 103 \msolarc. In order to increase further the dark matter resolution of our simulations 
we split the dark matter particles according to the prescription of \cite{Kitsionas_2002} and 
as described in \cite{Regan_2015}. We split particles centered on the position of the final collapse
as found from lower resolution simulations within a region with a comoving side length of 43.75 h$^{-1}$ kpc.
Each particle is split into 13 daughter particles resulting in a final high resolution region with a 
dark matter particle mass of $\sim$ 8 \msolarc. The particle splitting is done at a redshift of 40 well
before the collapse of the target halo. Convergence testing to study 
the impact of lower dark matter particle masses on the physical results was conducted and is discussed 
in \S \ref{Sec:DMConvergence}. \\
\indent The baryon resolution is set by the size of the grid cells, in the highest resolution region 
this corresponds to approximately 0.48  \kpch comoving (before adaptive refinement).  The  maximum 
refinement level for all of the simulations was set to 16. Refinement is triggered
in \enzo  when the refinement criteria are exceeded. The refinement criteria used in this work 
were based on three physical measurements: (1) The dark matter particle over-density, 
(2) The baryon over-density and (3) the Jeans length. The first two criteria introduce additional 
meshes when the over-density (${\Delta \rho \over \rho_{\rm{mean}}}$) of a grid cell with respect to 
the mean density exceeds 8.0 for baryons and/or DM. Furthermore, we set the 
\emph{MinimumMassForRefinementExponent} parameter to $-0.1$ making the simulation super-Lagrangian 
and therefore reducing the threshold for refinement as higher densities are reached. For the final 
criteria we set the number of cells per Jeans length to be 16 in these runs.

\subsection{Ray Tracing}
We enhanced the radiative transfer algorithm by upgrading 
the maximum \healpix \citep{Gorski_2005} level to 29. This allows for the ray tracing algorithm 
to penetrate even the densest grid structure created by the AMR framework within Enzo. Without 
including this modification the ray tracer is unable to properly resolve the most highly 
refined regions produced by Enzo's adaptive refinement mechanisms. The angular resolution of 
the ray tracing algorithm is determined by the number of pixels that the \healpix routines create 
as the rays propagate outwards. The angular resolution is given by 
\begin{equation}
\omega = \sqrt{4 \pi \over \rm{N_{pix}}}
\end{equation}
where $\omega$ is the angular resolution in steradians and $\rm{N_{pix}}$ is the number of pixels 
created by the \healpix solver. The number of pixels is given by 
\begin{equation}
\rm{N_{pix}} = 12 * 4^l
\end{equation}
where $l$ is the level of pixelisation. Using 64 bit numerical resolution $l$ can reach a maximum 
value of 29. As the rays propagate through the simulation they split when the associated solid angle
is greater than 1/$\kappa$ times the area of the cell that the ray is traversing where $\kappa$ is the number of 
rays per cell. In our simulations we set $\kappa$ to 5.1 (see \cite{WiseAbel_2011} for more details on this 
parameter) which is the default value. $l$ increases to allow the rays to split and the resolution of 
the ray tracer can always match the resolution of the AMR cells. \\
\indent The ray tracing solver in \texttt{Enzo}-3.0 is able to model the ionisation of H, He and He$^{+}$. 
It can also account for the photo-dissociation of \molH for photons with energies within the Lyman-Werner band. 
We have added further frequency channels to the ray tracing solver including $\rm{H^-}$ photo-detachment 
and $\rm{H_2^+}$ dissociation to complement the already existing algorithms. The ray tracer is therefore able 
to properly account for all of the relevant photo-ionisations and photo-dissociations relevant for studying
the direct collapse mechanism. Similarly to R14 we employ the self-shielding model of 
\cite{Wolcott-Green_2011} when calculating the \molH dissociation rate\footnote{We also ran simulations where we 
modelled the dissociation of \molH by a direct calculation of the optical depth (i.e. Dissociation Rate 
$\mathrm{\propto exp(-N_{H_2} \sigma_{H_2})}$), in this case we found very similar results with the 
temperature in the central regions of the collapsed halo being lower on average by about 300 K.}.

\begin{table*}
\centering
\caption{Radiation SED}
\begin{tabular}{ | l | c | c }
\hline 
\textbf{\em $\rm{Spectrum}$}
& \textbf{\em $\rm{Energy\ Bins\ (eV)}$} & \textbf{\em $\rm{Photon\ Fraction\ (PF)}$} \\
\hline 
\noindent \textit{No Extinction} \\ 
\ \ \ \ Blackbody (T = $10^4$ K) & 0.76,   8.0,    12.8 & 
                           0.9746, 0.0252, 0.0002 \\
\ \ \ \ Blackbody (T = $10^5$ K) & 0.76,   8.0,    12.8,   14.79,  20.46,   27.62, 60.0 & 
                           0.0795, 0.1440, 0.0745, 0.0741, 0.2484, 0.1124, 0.2671 \\
\ \ \ \ Stellar                  & 0.76,   8.0,    12.8,   14.79, 20.46,  27.62, 60.0 & 
                           0.4130, 0.3170. 0.1080, 0.414, 0.0399, 0.0324, 0.0278 \\
\textit{With Extinction}\\
\ \ \ \ Blackbody (T = $10^4$ K) & 0.76,   8.0,    12.8 & 
                           0.9746, 0.0252, 0.0002 \\
\ \ \ \ Blackbody (T = $10^5$ K) & 0.76,   8.0,    12.8,   14.79,  20.46,   27.62, 60.0 &
                           0.0795, 0.1440, 0.0745, 2.36e-07,  1.38e-03, 1.21e-02, 0.2122 \\

\ \ \ \ Stellar                  & 0.76,   8.0,    12.8,   14.79, 20.46,  27.62, 60.0 & 
                           0.4130, 0.3170, 0.1080, 1.32e-07, 2.23e-04,  3.49e-03,  2.26e-02 \\

\hline
\end{tabular}
\parbox[t]{0.9\textwidth}{\textit{Notes:} The energy bins and the fractional number of photons 
are given for each spectrum used in this study. The photon fractions are given for the cases
with and without extinction from the host galaxy ISM. The fractions without extinction are not 
used in this study but are included for reference for the reader. 
The blackbody spectrum with an effective temperature 
of T = $10^4$ K has no photons with energies greater than the 
ionisation threshold of hydrogen (13.6 eV) due to the exponential drop in the spectrum at 
energies greater than approximately 1 eV. In contrast the T = $10^5$ K blackbody spectrum peaks 
at energies greater than the ionisational threshold of hydrogen. The stellar spectrum has a 
more even distribution with a tilt towards energies in the infrared and optical. 
}

\label{Table:radiation_sed}
\end{table*}

\subsection{Chemical Modelling} \label{Sec:chemical_model}
We adopt here the 26 reaction network determined by \cite{Glover_2015a} as the most appropriate
network for solving the chemical equations required by the direct collapse model in a gas of 
primordial composition with no metal pollution. The network consists of ten individual species:
${\rm H}, {\rm H}^+, {\rm He}, {\rm He}^+,  {\rm He}^{++}, {\rm e}^-,$ 
$\rm{H_2}, \rm{H_2^+}\, \rm{H^-} \rm{and}\ \rm{HeH^+}$. Additionally, we included a further 7 
reactions which accounts for the recombinations (4) and photo-ionisations (3) of  ${\rm H}$, 
${\rm He}$, and ${\rm He}^{+}$ which occurs when the elements are photo-ionised due to photon energies greater 
than 13.6 eV, 25.4 eV and 54.4 eV, respectively. \\
\indent To implement the chemical network we have extensively modified the open source code 
\gracklec\footnote{https://grackle.readthedocs.org/}$^,$\footnote{Changeset: 88143fb25480} 
\citep{Enzo_2014, Kim_2014}. \grackle self-consistently 
solves the 33 set reaction network including photo-ionisations. The network includes the most 
up-to-date rates as described in \citet{GloverJappsen_2007, GloverAbel_2008, GloverSavin_2009,
Coppola_2011, Coppola_2012,  Glover_2015a, Glover_2015b,  Latif_2015}. The reaction network
is described in full in Table \ref{Table:reaction_network}. 
The gas is allowed to cool radiatively during the simulation and this is also accounted for 
using the \grackle module. Here the rates have again been updated to account for recent updates 
in the literature \citep{Glover_2015a}. The cooling mechanisms included in the model are collisional
excitation cooling, collisional ionisation cooling, recombination cooling, bremsstrahlung and 
Compton cooling off the CMB. 

\subsection {Models} \label{Sec:Models}
For this study we analyse a single halo. The halo studied is identical to one used in 
\cite{Regan_2015} with the initial conditions created with the \texttt{MUSIC} code. 
The central idea is to place a radiating source close to a collapsing halo and investigate the effect 
of a realistic radiation field on the collapse of the halo and to determine the viability of the 
direct collapse method. The idea that close-by neighbours are required for direct collapse has previously 
been studied analytically by \cite{Dijkstra_2008, Dijkstra_2014} and more recently using synchronised halo pairs 
by \cite{Visbal_2014b}. For each simulation we vary the source characteristics (SED) and the 
distance of the source to the maximum density point. In each case the simulation is 
initialised at z = 99 and evolved until a redshift of 40.0. At this point, the pre-galactic cloud 
has started to assemble but has not yet reached a mass that supports PopIII star formation.
Continuing to evolve the simulation at this point would result in the formation of a Pop III star 
at a redshift of $z \approx 33$ (see \textit{Ctrl} simulation in Table \ref{Table:radiation_particle}). 
We now select the point of maximum density at z = 40 and 
place a single \textit{radiation particle} at a distance of 0.5 kpc, 1 kpc and 2 kpc resulting in 
three different simulations. Further models are defined for the stellar spectrum case with initial 
distances of 4.0 kpc, 8.0 kpc and 12.0 kpc. The flux from the source is identical in each case and only the 
spectral energy distribution and distance from the point of maximum density changes in each case. The 
specification of the \textit{radiation particles} are listed in Table 
\ref{Table:radiation_particle} resulting in a total of 13 different models.

\subsection{Dark Matter Convergence} \label{Sec:DMConvergence}
In order to evaluate the extent to which convergence is achieved in our simulations
we follow the study undertaken by \cite{Regan_2015}. Their rule of thumb for dark matter 
resolution in simulations of high redshift collapse states that 
\begin{equation}
{M_{\rm{core}} \over M_{\rm{DM,part}}} > 100.0
\end{equation}
where M$_{\rm{core}}$ is the baryonic mass within the core\footnote{The core of the halo is defined 
at the point where the baryonic mass exceeds the dark matter mass. This fluctuates between 
approximately 1 pc and 5 pc across the simulations. We therefore choose 1 pc to define the radius 
of the core of the halo in all cases for consistency.} of the halo and M$_{\rm{DM,part}}$ is the 
dark matter particle mass (equivalent to the dark matter mass resolution). In Table 
\ref{Table:radiation_particle} the core mass of the 
halo is displayed in the final column. As noted above our dark matter mass resolution for all of 
our simulations is $\sim 8$ \msolarc. We have performed simulations at lower resolution and found that 
the resulting differences in the thermal history of the model are negligible and are confident we
have achieved convergence for this resolution scale. 
\begin{figure*}
  \centering 
  \begin{minipage}{175mm}      \begin{center}
      \centerline{
        \includegraphics[width=9cm]{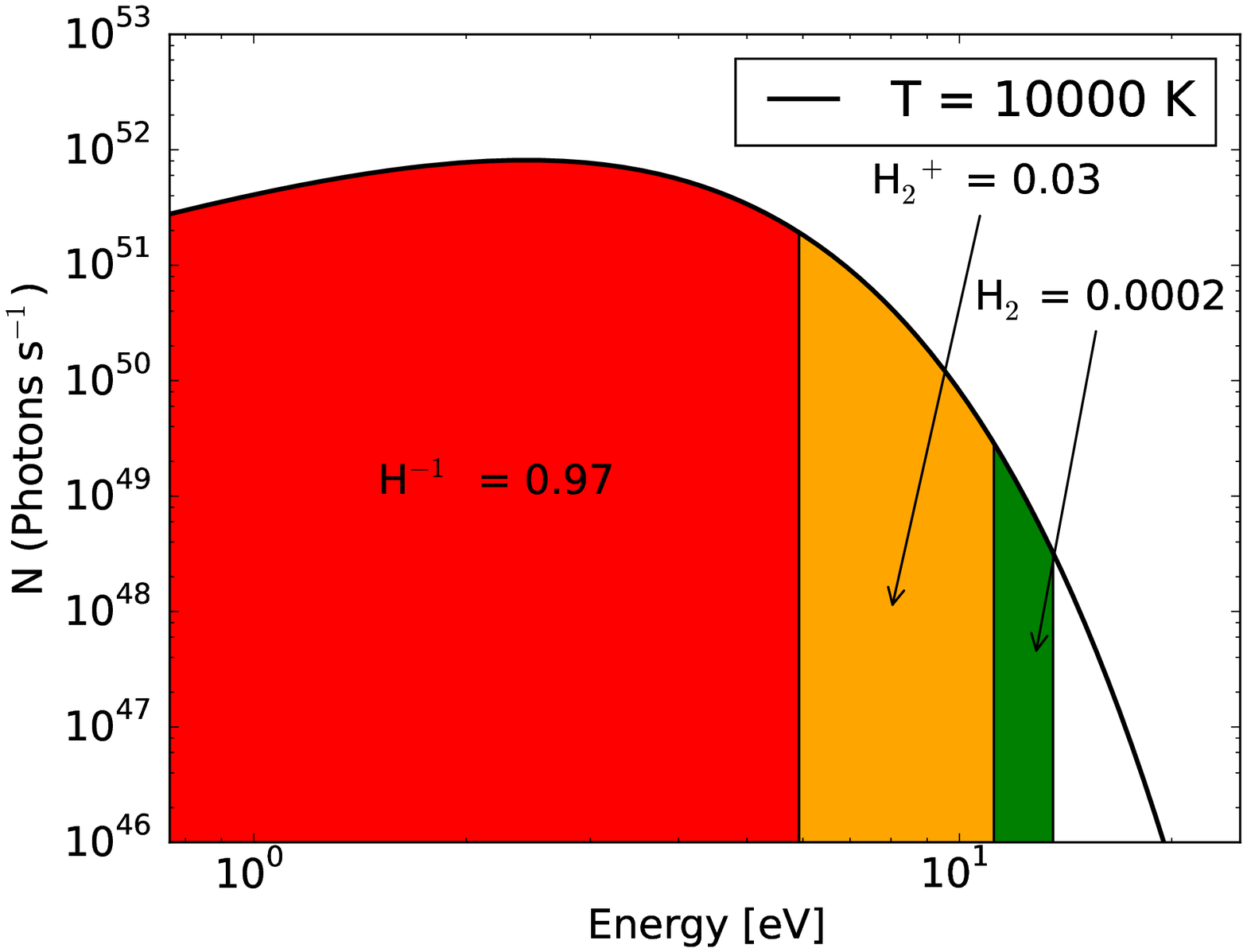}
        \includegraphics[width=9cm]{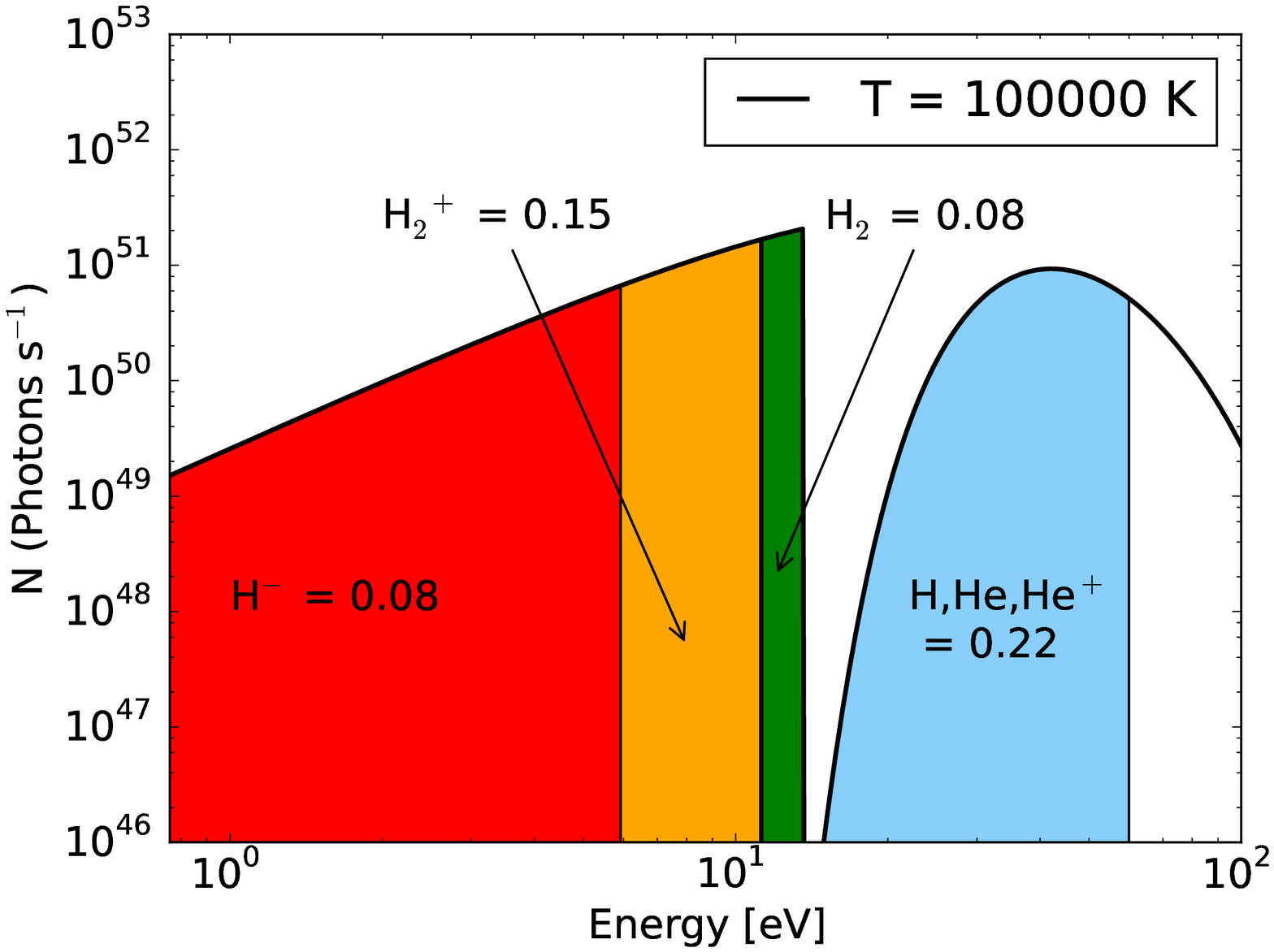}}
        
        \caption[]
        {\label{BlackbodySED}
        The left panel shows the blackbody spectrum with an effective temperature of 
        $10^4$ K. The fractional number of photons which effects each species is shown (or pointed to 
        with an arrow) for each species. The right hand panel shows the blackbody spectrum when the 
        effective temperature is $10^5$ K. In this case the spectrum includes photons which can 
        ionise hydrogen and helium. The $10^5$ K spectrum also includes the extinction factor due to 
        absorption from the ISM which we include in our models. The absorption of ionising photons 
        causes the large gap in the spectrum around 13.6 eV. 
        }
      \end{center} \end{minipage}
  \end{figure*}

\subsection{Radiation Source}
The radiation source is a point particle. It is massless and is fixed in comoving space. 
The physical distance between the source and the collapsing halo therefore inevitably increases 
due to the expansion of the universe at this redshift. 
The source of radiation is placed at a distance of between 0.5 kpc and 12 kpc, depending on the given 
model being tested, from the point of maximum density at a redshift of 40. In each case, 
we use a luminosity of $1.2 \times 10^{52}$ photons per second (above the H$^-$ photo-detachment 
energy of 0.76 eV) that originates from a galaxy with a stellar mass of $10^3$ \msolar at $z = 40$. 
The galaxy has a  specific star formation rate (SFR) of $\rm sSFR=40 \ \rm Gyr^{-1}$ resulting in a stellar mass of 
$10^5$ \msolar at $z = 20$.  The stellar mass at $z = 20$ and the specific SFR are 
consistent with the largest galaxies prior to reionisation in the Renaissance Simulations of 
\cite{Chen_2014}.  We then calculate its spectrum with the \cite{Bruzual_2003} models with a 
metallicity of $10^{-2}$ \zsolar and compute the photon luminosity from it. The spectrum does not 
include emission from the nebular component and is solely due to stellar emission.\\ 
\indent As stated in the Introduction J$_{21}$ is the standard unit used to measure radiation 
(background) intensities. It is a measure of the intrinsic brightness or intensity of a source which 
is assumed to be constant at all points in space. We therefore also quote this value to be consistent 
with the literature noting however that in all cases our radiation is from a single direction and not 
constant at all points. To calculate the intensity, $J$, in units of J$_{21}$ we sum the contributions to 
$J$ from each energy bin used in our model and normalise $J$ at the hydrogen ionisation edge as follows:

\begin{align}
J^\prime & = \sum_{E, i}  {k_i  E \over 4 \pi^2 \sigma_i(E)} \\
J  & = {J^\prime \over \nu_{H} J_{21}}
\end{align}
where $J^\prime$ is the sum of the intensities for each species, $i$, over all energy bins, 
$E$. Here $k_i$ is the number of photo-ionisations (or dissociations) per second for species $i$, 
$\sigma_i(E)$ is the cross section for species $i$ at energy $E$. Finally, $\nu_{\rm{H}}$ is the 
frequency at the hydrogen ionisation edge. The extra factor 
of $\pi$ in the denominator accounts for the solid angle. $J$ is now the intensity of the radiation 
background in units of J$_{21}$. Individual contributions to the intensity are summed and normalised 
at the hydrogen edge, the normalisation of individual contributions follows the same procedure 
described elsewhere in the literature  \citep[e.g.][]{Haiman_2000}. By summing over the individual 
contributions to the intensity and normalising at the hydrogen edge we are able to display a single, 
well defined, value for the intensity at all points as a function of distance from our source. 
Note that this definition of the mean intensity differs somewhat from those used in previous studies of 
DCBH formation where a background intensity is used. Care should be taken when 
comparing our values of J with the values quoted in these studies as ours are due to an anisotropic 
multi-frequency radiation source.\\
\indent In this study we make use of three different methods to control the spectral energy 
distribution (SED) of the source. We use a black body spectrum with an effective temperature of
T = $10^4$ K and one with T = $10^5$ K consistent with previous studies of dissociating 
\molH \citep{Shang_2010, Latif_2015}. In order to account for energy in different radiation bands 
we use a seven bin model to probe the effect of radiation with different energies. As such we 
select radiative transfer bins with energy of 0.76, 8.0, 12.8, 14.79, 20.46, 27.62 and 60 eV. 
The final energy bin is artificially cut just above the double ionisation threshold of Helium (54.42 eV).
We will investigate the effects of X-ray radiation, with energies greater than 60 eV, in a follow-up paper. 
The first three energy bins are weighted by the cross section peaks for $\rm{H^-}, \rm{H_2^+}\ and \ 
\rm{H_2} $ photo detachment/dissociation respectively. The final four energy bins are determined 
using the \texttt{sedop} code developed by \cite{Mirocha_2012} which determines the optimum 
number of energy bins needed to accurately model radiation with energy above the ionisation 
threshold of hydrogen. For computational reasons we do not include more than seven energy bins in 
this study as the ray tracer scales with the number of energy bins used and the simulations would quickly 
become computationally too demanding. \\
\indent Furthermore, we do not attempt to take into account any sources of background radiation in 
our model. At these redshifts (z $\gtrsim$ 20) the background flux in the Lyman-Werner is likely to be very 
small \citep{Dijkstra_2008, Johnson_2008} and instead we require a nearby source to provide a strong, local, 
dissociating flux.

\begin{figure*}
  \centering 
  \begin{minipage}{175mm}      \begin{center}
      \centerline{
        \includegraphics[width=9cm]{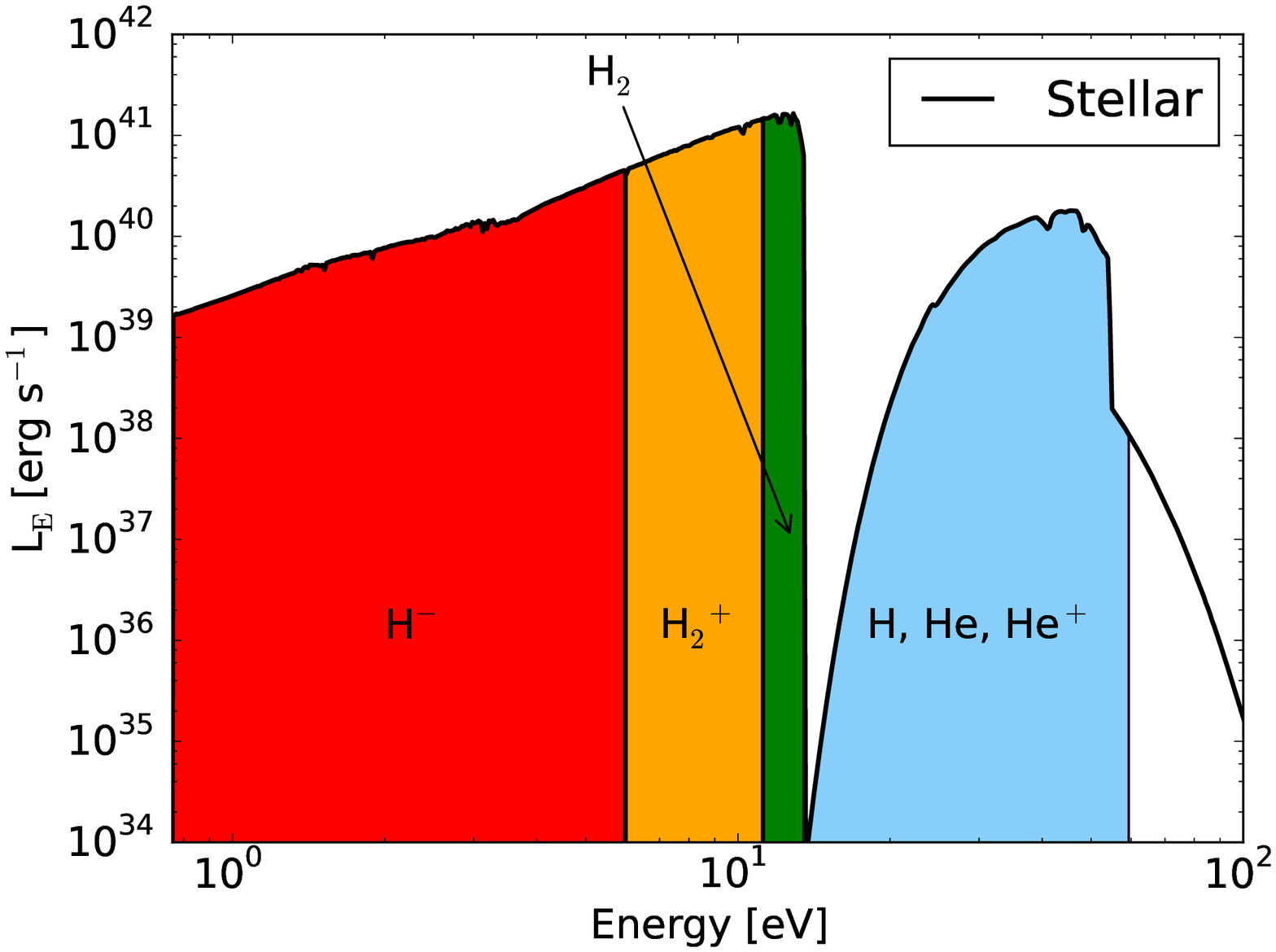}
        \includegraphics[width=9cm]{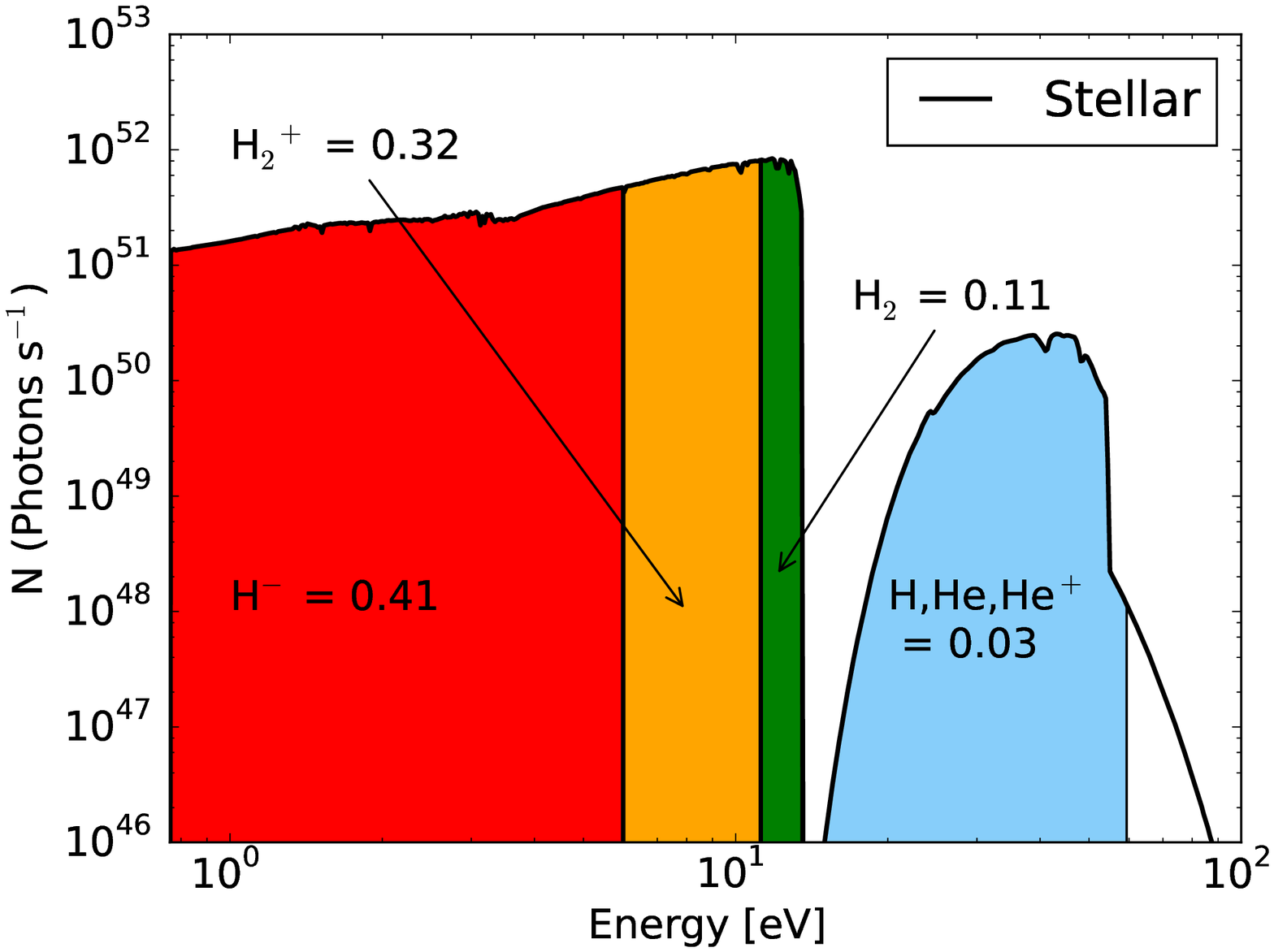}}
        \caption[]
        {\label{StellarSED}
        The left panel shows the luminosity from a stellar spectrum extracted from the 
        Renaissance Simulation of \cite{Chen_2014}. The total stellar mass of the 
        spectrum is $1 \times 10^5$ \msolar at z = 20. As with the blackbody spectrum of $10^5$ K
        we employ an extinction factor for photons with energy greater than 13.6 eV and a 
        cutoff for photons greater than 60 eV. In the right hand panel we plot the 
        same spectrum with the photon luminosity (in units of photons per second) on the y-axis.
        The fraction of photons in each energy band is indicated. For the case of the 
        stellar spectrum most of the photons are lower energy photons with energies less than 
        13.6 eV. 
        }
      \end{center} \end{minipage}
  \end{figure*}


\begin{figure*}
  \centering 
  \begin{minipage}{175mm}      \begin{center}
      \centerline{
        \includegraphics[width=17cm]{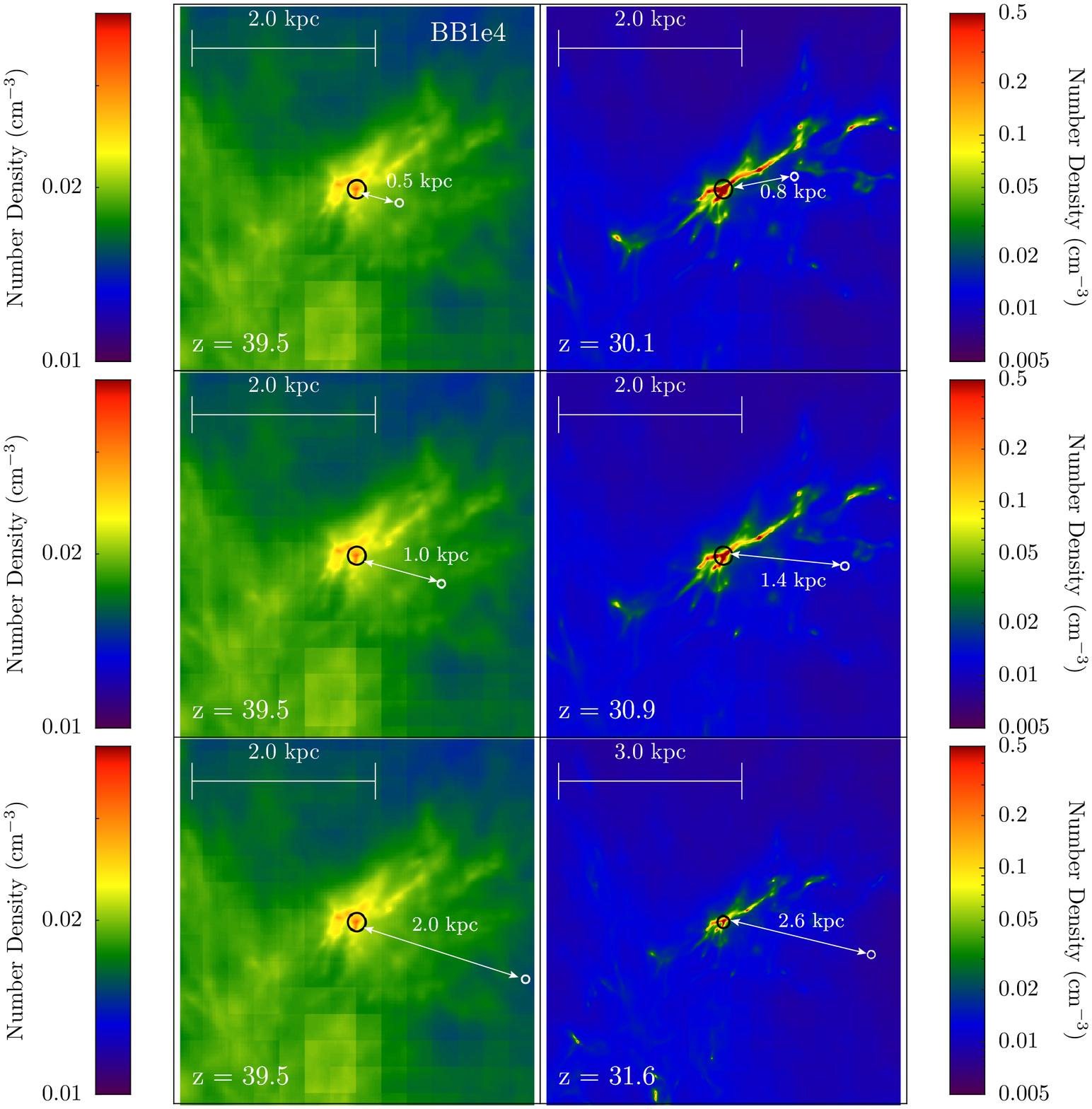}}
        \caption[]
        {\label{HaloB_Projections_BB1e4}
          The Figure contains projections for the three models run with a blackbody 
          spectrum with T = $10^4$ K (BB1e4). In the top row we show the model where the source is set 
          at an initial distance of 0.5 kpc from the target halo, in the middle row the source 
          is placed at a distance of 1.0 kpc while in the bottom row the source is placed at a distance 
          of 2.0 kpc. In the left hand column we show the setup a short time after the source is 
          switched on, at a redshift of 39.5. In the right hand column we show the final output 
          of the simulation. The source is marked with a white circle. The point of maximum density
          is identified with a black circle. The larger distances between the source and the point of
          maximum density shown in the right hand column is due to the expansion of the universe over the
          given redshift interval.
          
        }
      \end{center} \end{minipage}
  \end{figure*}

\begin{figure*}
  \centering 
  \begin{minipage}{175mm}      \begin{center}
      \centerline{
        \includegraphics[width=17cm]{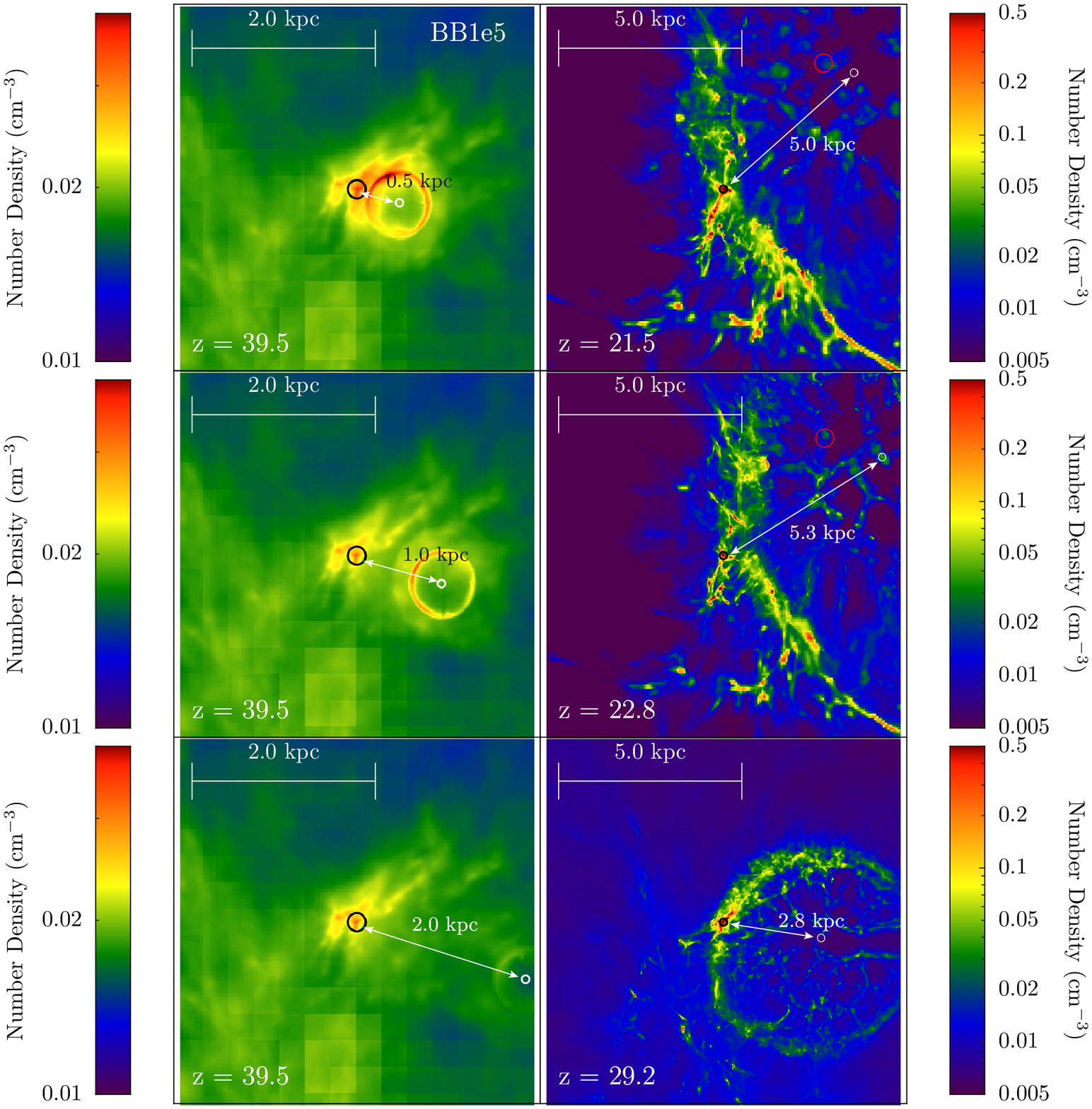}}
        \caption[]
        {\label{HaloB_Projections_BB1e5}
          The Figure contains projections for the three models run with a blackbody 
          spectrum with T = $10^5$ K (BB1e5). In the top row we show the model where the source is set 
          at an initial distance of 0.5 kpc from the target halo, in the middle row the source 
          is placed at a distance of 1.0 kpc while in the bottom row the source is placed at a distance 
          of 2.0 kpc. In the left hand column we show the setup a short time after the source is 
          switched on, at a redshift of 39.5. In the right hand column we show the final output 
          of the simulation. The source is marked with a white circle. The point of maximum density
          is identified with a black circle.  The larger distances between the 
          source and the point of maximum density shown in the right hand column is due to the expansion 
          of the universe over the
          given redshift interval. The red circle in the top right panel and the middle right panel 
          indicates the approximate location of the original target halo (which is completely disrupted). 
          
        }
      \end{center} \end{minipage}
  \end{figure*}


\begin{figure*}
  \centering 
  \begin{minipage}{175mm}      \begin{center}
      \centerline{
        \includegraphics[width=17cm]{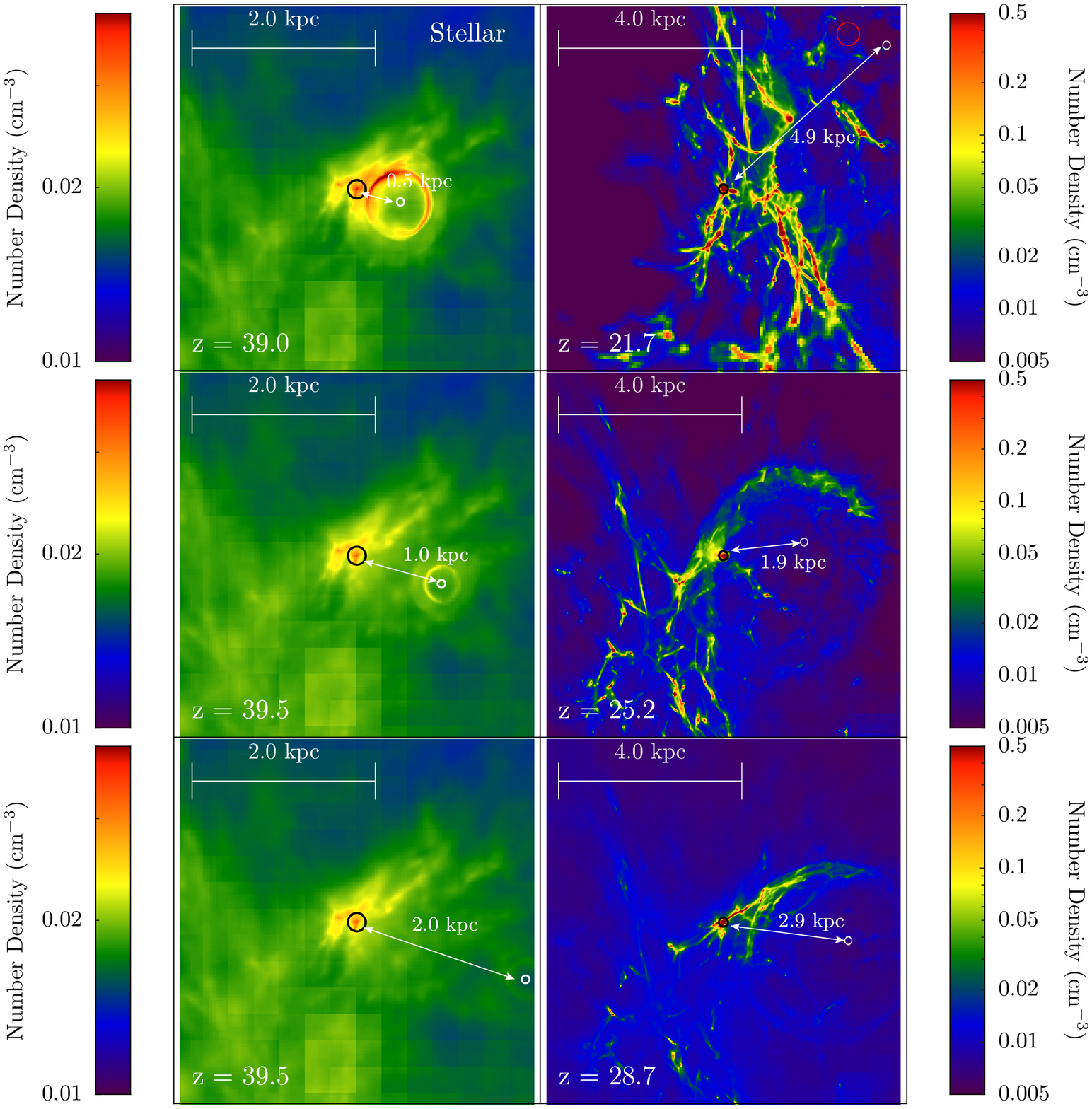}}
        \caption[]
        {\label{HaloB_Projections_SED}
          The Figure contains projections for the three models run with a realistic stellar SED. 
          In the top row we show the model where the source is set 
          at an initial distance of 0.5 kpc from the target halo, in the middle row the source 
          is placed at a distance of 1.0 kpc while in the bottom row the source is placed at a distance 
          of 2.0 kpc. In the left hand column we show the setup a short time after the source is 
          switched on, at a redshift of 39.5. In the right hand column we show the final output 
          of the simulation. The source is marked with a white circle. The point of maximum density
          is identified with a black circle. The larger distances between the source and the point of
          maximum density shown in the right hand column is due to the expansion of the universe over the
          given redshift interval. The red circle in the top right panel
          indicates the approximate location of the original target halo (which is completely disrupted).

        }
      \end{center} \end{minipage}
  \end{figure*}


\subsection{Sampling the Energy Spectrum}
The energies and fractional number of photons in each energy bin for all three models are shown in 
Table \ref{Table:radiation_sed}. During test runs we found that it was necessary 
to include a method to model the extinction properties of the interstellar medium (ISM) of the source galaxy. 
Without extinction the hydrogen ionising radiation created an unrealistically large HII region around the 
radiation source. In order to increase the realism of our simulations we instead adopted a simple model of 
ISM extinction. The model convolves the spectral energies from our three fiducial spectra with a simple 
modelling of the optical depth to ionising radiation as follows:

\begin{equation}
\rm{PF_{ext}(E)} = PF(E) \times \rm{exp}(-\sigma(E_{ph}) \times N(HI)_{avg})
\end{equation}
where $\rm PF(E)$ is the photon fraction at the energy, $\rm E$,  $\rm PF_{\rm{ext}}(E)$ is the 
photon fraction when the extinction is accounted for, $\sigma(\rm E)$ is the cross 
section of hydrogen at that energy and $\rm N(HI)_{\rm{avg}}$ is the column density of
hydrogen averaged over the source galaxy. For our model we choose an average value of 
N(HI)$_{\rm{avg}}$ of $2.5 \times 10^{18} \rm{cm^{-2}}$ consistent with the results from the simulations of 
\cite{Wise_2009}. Physically this is motivated by the fact that low density channels of 
neutral hydrogen allow for the escape of ionising radiation between approximately 13.6 and 50 eV
from the radiating galaxy. These channels are somewhat transient and evolve over time \citep{Wise_2014} meaning 
that over a sufficient amount of time (approximately 80 Myrs) the halo receiving the flux is 
swept over by ionising radiation in these bands rather than being illuminated constantly. Our 
ISM modelling is an attempt to take this effect into account. In Table \ref{Table:radiation_sed} we 
therefore include the photon fraction both with and without the extinction factor so that the 
reader can easily see the differences. The extinction factor is set to 1.0 for energy below the 
ionisation threshold of neutral hydrogen, thus having no effect in that case. The mean free path 
of photons below the ionisation threshold of hydrogen is comparatively long and is not included in our model.
Strong internal Lyman-Werner flux will dissociate most of the \molH in the source galaxy with the exception of 
some molecular clouds, which have a small geometric cross-section and can be safely ignored.
We only employ the photon fractions which include the extinction factor in our production runs 
(i.e. the bottom three lines).\\  
\indent The spectrum in the case of the T = $10^4$ K blackbody is strongly 
tilted towards radiation with energy in the optical and infrared. The exponential fall off in the 
number of photons with energies greater than a few electron volts means that there are virtually no 
photons capable of ionising hydrogen with this spectrum. For the blackbody spectrum with 
T = $10^5$ K the peak in the SED is shifted towards higher energies and in this case the 
photons are capable of ionising hydrogen and helium. The spectrum based on a stellar
SED is more evenly distributed but with a clear tilt towards lower energy photons. A 
plot of each spectrum is shown in Figures \ref{BlackbodySED} and \ref{StellarSED}. \\
\indent In the left hand panel of Figure \ref{BlackbodySED} we have plotted the spectrum for the 
$10^4$ K blackbody spectrum. Energy in eV is plotted on the x-axis while the number of photons emitted 
per second is plotted on the y-axis. For the case of the $10^4$ K spectrum the peak in the SED occurs at
between 1 and a few eV in the infrared part of the spectrum and most photons are emitted with this energy. 
No photons with energies greater than the ionisation threshold of hydrogen are emitted in our model for this
spectrum. In the right hand panel of Figure \ref{BlackbodySED} we have made the same plot for the 
$10^5$ K blackbody spectrum. In this case the effect of the extinction factor is clearly evident. The 
extinction manifests itself as a sharp drop in the photon count at energies greater than 13.6 eV before 
eventually recovering as the cross section to hydrogen falls off to higher energies. The gap in the spectrum
accounts for absorptions by the ISM. It should also be noted that in this case the sum of the fractions does 
not equate to unity (0.08 + 0.15 + 0.08 + 0.22 = 0.53 $\ne$ 1.0). These fractions are passed to the ray tracer
via a parameter file and as a result the number of hydrogen ionising photons is strongly reduced in cases where 
extinction is included. \\
\indent In the left hand panel of Figure \ref{StellarSED} we show the SED for the case of the stellar spectrum and 
in the right panel we show the photon number count versus energy. The stellar spectrum peaks in the UV with 
a significant fraction of the photons contributing to the direct dissociation of \molH via the Solomon process.
As well as this, there is a substantial fraction of photons which contribute towards photo-detaching the two 
intermediary 
species $\rm{H^-}\ and\ \rm{H_2^+}$. The fraction of hydrogen ionising photons in the Stellar SED is significantly 
lower than the $10^5$ K spectrum but nonetheless non-zero and so a HII region will be forming around the source 
and over time will expand. The values of the individual energies of each energy bin are available in 
Table \ref{Table:radiation_sed}.
\begin{figure*}
  \centering 
  \begin{minipage}{175mm}      \begin{center}
      \centerline{
        \includegraphics[width=18cm]{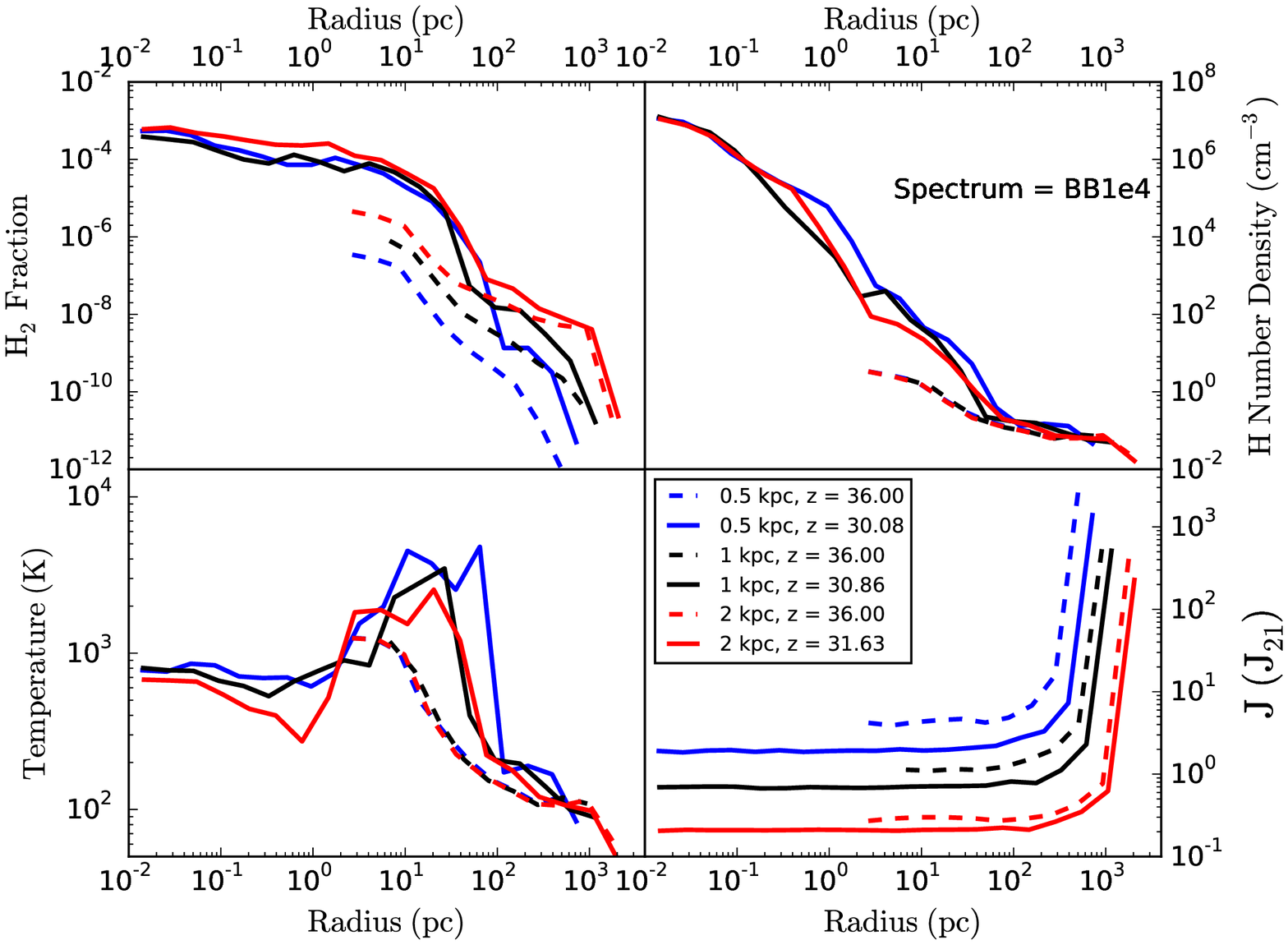}}
        \caption[]
        {\label{HaloB:MultiPlot_BB1e4}
          \textit{BB1e4}: Ray profiles for when the halo is exposed to a radiation source with a 
          10000 K blackbody spectrum. In each panel the quantity displayed is a function of radius 
          along 1000 sight lines from the radiation source to the central density at each redshift.
          The top left panel shows the \molH fraction, the top right panel shows the hydrogen 
          number density, the bottom right panel shows the value of the intensity, $\rm{J}$, in the usual units
          of \J, the bottom left panel shows the temperature. 
          }
      \end{center} \end{minipage}
  \end{figure*}

\begin{figure*}
  \centering 
  \begin{minipage}{175mm}      \begin{center}
      \centerline{
        \includegraphics[width=18cm]{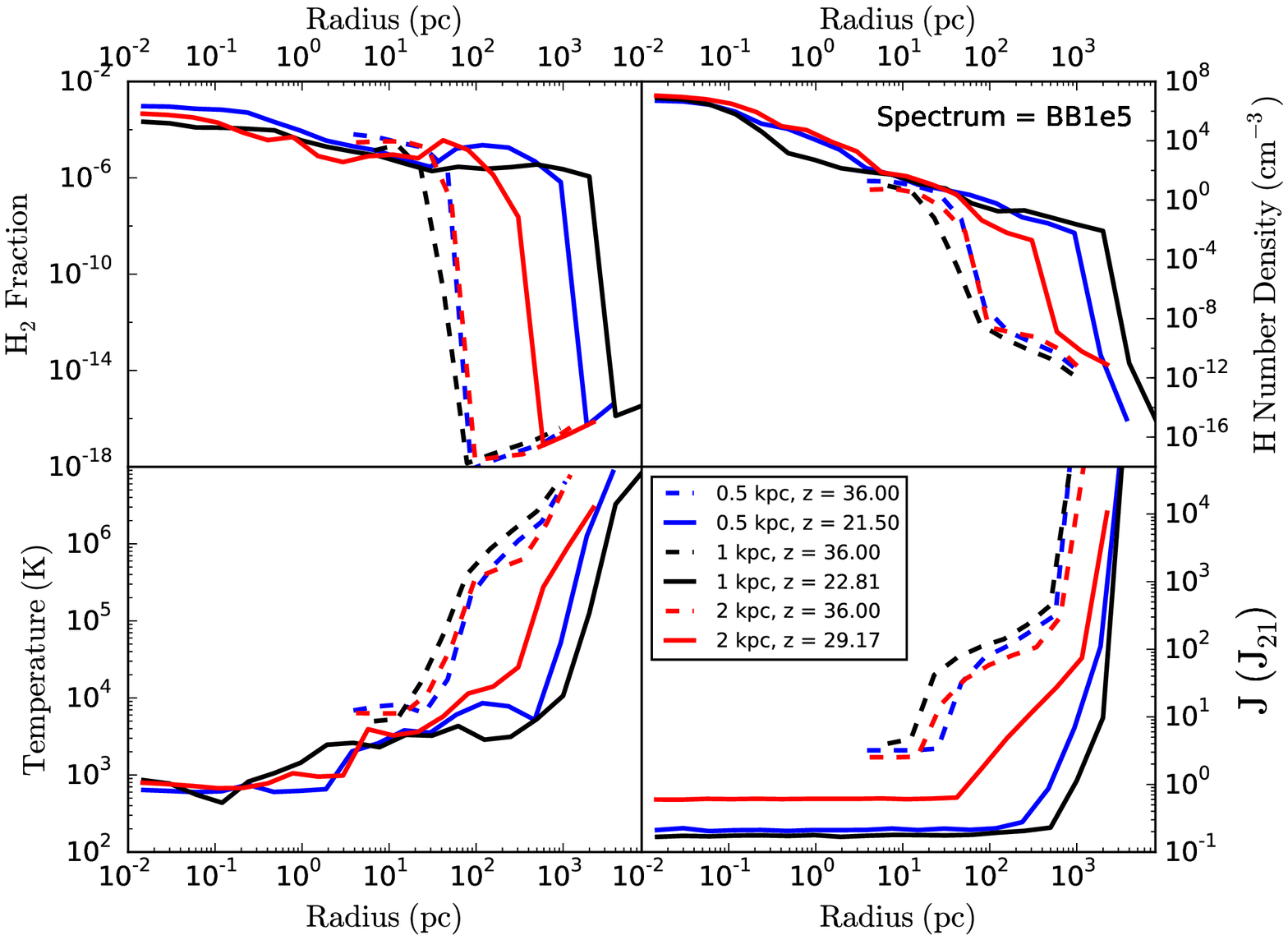}}
        \caption[]
        {\label{HaloB:MultiPlot_BB1e5}
          \textit{BB1e5}:  Ray profiles for when the target halo is exposed to a radiation source with a 
          100000 K blackbody spectrum. In each panel the quantity displayed is a function of radius 
          along 1000 sight lines from the radiation source to the central density at each redshift.
          The top left panel shows the \molH fraction, the top right panel shows the hydrogen 
          number density, the bottom right panel shows the value of the intensity, $\rm{J}$, in the usual units
          of \J, the bottom left panel shows the temperature. 
          }
      \end{center} \end{minipage}
  \end{figure*}

\section{Results} \label{Sec:Results}
Each of the models described in Table \ref{Table:radiation_particle} result in a qualitatively 
different result. The difference in the initial distance to the source and the SED of that source mean that 
the collapse of the nearby halo is either significantly delayed or else the collapse is prevented 
entirely and a different halo collapses (i.e. one that is further from the source). We will now discuss the 
impact of each spectrum type (BB1e4, BB1e5 and Stellar SED) on the target halo. We will begin by 
looking at the visual impact of the radiation fields before looking more quantitatively at the 
impact of the radiation.\\

\subsection{Visual Inspections}
\subsubsection{Blackbody with T = $10^4$ K (BB1e4)}
In Figures \ref{HaloB_Projections_BB1e4}, \ref{HaloB_Projections_BB1e5} and  \ref{HaloB_Projections_SED} we 
show projections for all  models where the initial distance is 0.5 kpc, 1.0 kpc or 2.0 kpc. We mark the source 
with a white circle, the point of maximum density with a black circle and the target halo with a red circle. The
target halo is only marked in the cases where it does not overlap with the point of maximum density (i.e. in 
simulations where the radiation from the source disrupts entirely the target halo). The distance from the source 
to the collapse halo is identified in each case. 
As we will see the target halo is not always the halo which collapses first. Looking first at Figure 
\ref{HaloB_Projections_BB1e4} where we plot the visualisation for the realisations with a blackbody spectrum of 
T = $10^4$ K (BB1e4) the target halo and collapse halo are the same halo. The distances from the source to 
the target halo 
changes with time due to cosmic expansion and the evolution of the system (the source is fixed in comoving space). 
The collapse shifts to lower redshifts as the source is brought closer to the target halo due to the increased 
dissociation of \molH from the radiation source. However, in each case the overall morphology of the 
system remains unchanged. Note also that in each of these cases the final virial temperature of the halo remains
well below the atomic cooling threshold indicating that the dominant coolant remains \molH in all of the cases.\\ 
\subsubsection{Blackbody with T = $10^5$ K (BB1e5)}
\indent In Figure \ref{HaloB_Projections_BB1e5} the models for a 
source with a blackbody spectrum of T = $10^5$ K (BB1e5) is plotted. In this case the target halo and the 
collapse halo do not match when the source is placed at a distance of 1 kpc or 0.5 kpc. Cosmic expansion 
will account for an approximate doubling of the physical distance between the source and the target halo from 
z = 40 to z = 20 while in the 0.5 kpc and 1.0 kpc cases the distance between the source and the target 
halo is significantly greater by a factor of 2. We have marked the approximate location of the target 
halo's position in these projections using a red circle. The radiation from the source in this case has 
completely disrupted the target halo. The ionising 
radiation with energy greater than 13.6 eV has prevented the halo from cooling and has also heated the gas 
further reducing its ability to cool. The result is that a halo at larger distances from the source has 
collapsed. When the source is placed further from the target halo - at a distance of 2 kpc - the target halo 
and the collapse halo remain the same and target halo collapses at a redshift of z = 29.2. In the 05-T5 and 
1-T5 cases the collapse occurs at the edge of the HII region. Within this shell, collapse is prevented by the 
ionising radiation field which prevents efficient cooling of the gas. On the HII shell the gas is able to cool 
and condense and in both cases a collapse occurs. The 2-T5 model is not hampered in the same way. In this
case the HII region again forms and this is seen quite clearly in the bottom row of Figure 
\ref{HaloB_Projections_BB1e5}. In this case the HII region is unable to envelop the target halo before it 
collapses. From the projection it can be clearly seen that the target halo is collapsing just outside the 
HII region and so it is able to escape the damaging effects of the ionising radiation on its ability to cool. \\
\subsubsection{Stellar Spectrum}
\indent In Figure \ref{HaloB_Projections_SED} the models for the source with a stellar like 
spectrum are shown. Again the column on the left shows the source and target halo shortly after the source is 
turned on. The column on the right shows the collapsed halo at the final output time. The outline of the 
HII region is clearly visible in the middle and bottom rows of the plots. In the top row the initial distance 
to the target halo is set to be 0.5 kpc. The ionising radiation overwhelms the target halo in this instance 
and that halo is then not the first halo to collapse. Rather in this case a halo at 5.0 kpc from the source 
collapses first - outside the sphere of influence of the ionising radiation. In this case the halo again 
collapses just outside the HII region where dense gas is able to cool effectively. Again we have marked the 
position of the original target halo with a red circle. 
In the middle row the target halo does undergo collapse. The distance to the target halo at the time of collapse 
is 1.9 kpc (as expected) and the halo lies close to the edge of the HII region. In the bottom row the target 
halo is again the collapse halo with a distance from the source of 2.9 kpc. The halo lies just outside the 
HII region in this case and so again is able to collapse. We will now examine the collapse of each halo more 
quantitatively. 
 
\begin{figure*}
  \centering 
  \begin{minipage}{175mm}      \begin{center}
      \centerline{
        \includegraphics[width=18cm]{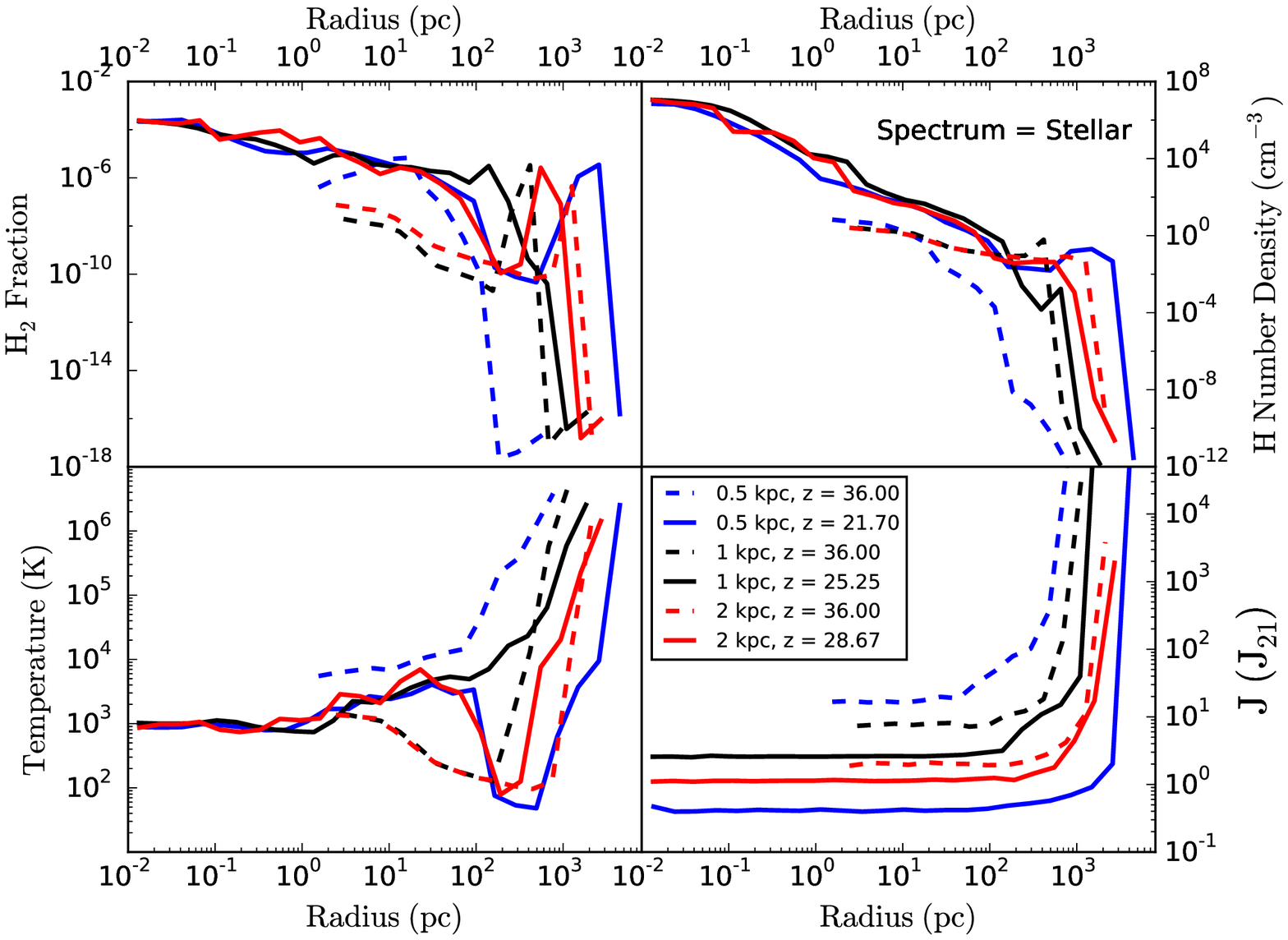}}
        \caption[]
        {\label{HaloB:MultiPlot_SED}
          \textit{Stellar SED}:  Ray profiles for when the target halo is exposed to a radiation source 
          based on a realistic Stellar SED. In each panel the quantity displayed is a function of 
          radius along 1000 sight lines from the radiation source to the central density at each 
          redshift. The top left panel shows the \molH fraction, the top right panel shows the 
          hydrogen number density, the bottom right panel shows the value of the intensity, $\rm{J}$, 
          in the usual units of \J, the bottom left panel shows the temperature. 
          }
      \end{center} \end{minipage}
  \end{figure*}
\begin{figure*}
  \centering 
  \begin{minipage}{175mm}      \begin{center}
      \centerline{
        \includegraphics[width=18cm]{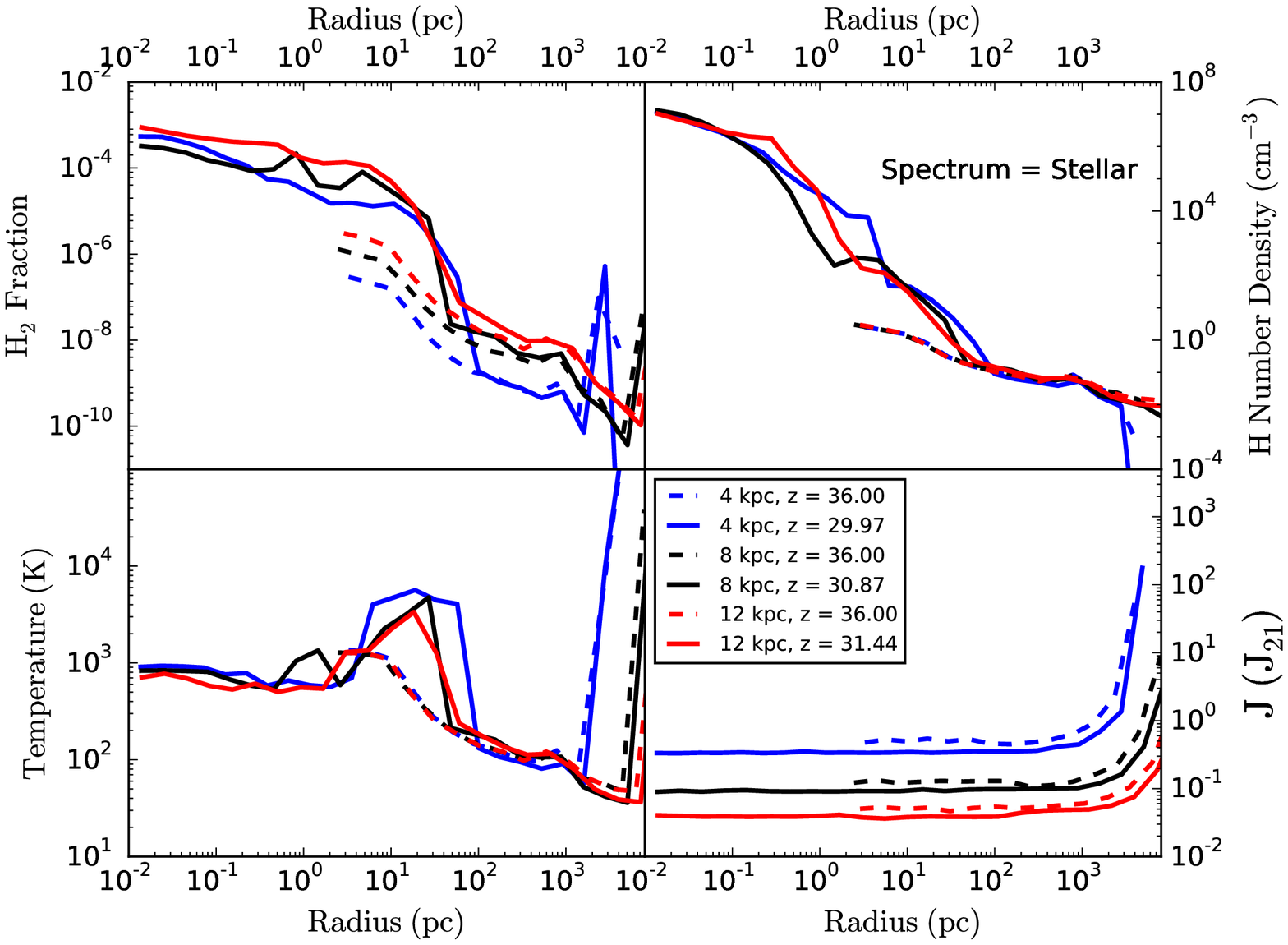}}
        \caption[]
        {\label{HaloB:IncreasedDistance}
         \textit{Stellar SED at increased distances}: Ray profiles showing the effect of radiation at greater 
         distances. As we increase the distance the temperature within the virial radius ($\sim$ 10 - 100 pc) 
         systematically decreases for the models as the initial separation is varied from 4 kpc to 12 kpc. 
         As we move to 12 kpc the temperature variation is clearly smaller and saturation is setting in. 
         In all three cases the primary coolant is \molH - the \molH rises above a mass fraction 10$^{-6}$ 
         at approximately 75 pc giving virial temperatures of $\lesssim$ 3000 K is each case. 
          }
      \end{center} \end{minipage}
  \end{figure*}
\subsection{Ray Profiling}

\subsubsection{Blackbody with T = $10^4$ K}
In Figures \ref{HaloB:MultiPlot_BB1e4}, \ref{HaloB:MultiPlot_BB1e5} and \ref{HaloB:MultiPlot_SED} we have plotted 
ray profiles from the source to the collapsing halo. The shown profiles are averaged over 1000 sight lines.
The lines all start from the source and travel to an area surrounding 
the collapsing halo. The first sight-line always connects the source and the point of maximum density. Each 
subsequent sight-line is given a small (randomly generated) angular offset so that it traverses a slightly 
different path to the central region thus defining a circular region around the central region from 
which the sight-lines are extracted.  
In Figure \ref{HaloB:MultiPlot_BB1e4} we show the ray profiles for all of the models
for the blackbody source with a temperature of T = 10$^4$ K. For each source with a different initial distance 
we show the ray profile at z = 36.0 (before any collapse) and at the collapse redshift. \\
\indent Given that the flux received at the target halo is significantly different at the time of collapse 
(see bottom right panel of Figure \ref{HaloB:MultiPlot_BB1e4}) the overall characteristics of the halos are 
still quite similar. In all cases the temperature at the centre of the halo converges to approximately 
T = 700 K and the \molH fraction is also approximately the same in each case. Some differences emerge at a 
distance of greater than approximately 1 pc. Within 1 pc self-shielding of \molH takes over meaning 
that within this radius the dissociating radiation has little effect and hence we see similar characteristics
across all three models within this radius. At a radius greater than 1 pc the effects of the dissociating 
radius are more obvious and we see that the model with a initial separation of 2 kpc has a 
systematically reduced temperature (red line) at almost all scales.  This is due to the enhanced \molH fraction 
compared to the other models where the source is placed closer to the target halo. The dissociating 
\molH radiation is unable to destroy the \molH as efficiently and the cooling due to \molH is more effective. For 
cases where the source has an initial separation of 0.5 kpc or 1 kpc the quantitative and qualitative 
differences are rather small. The flux reaching the central object in both of these cases is approximately 2.0  
J$_{21}$ for an initial separation of 0.5 kpc and 1 J$_{21}$ for an initial separation of 1.0 kpc.
\begin{figure*}
  \centering 
  \begin{minipage}{175mm}      \begin{center}
      \centerline{
        \includegraphics[width=9cm]{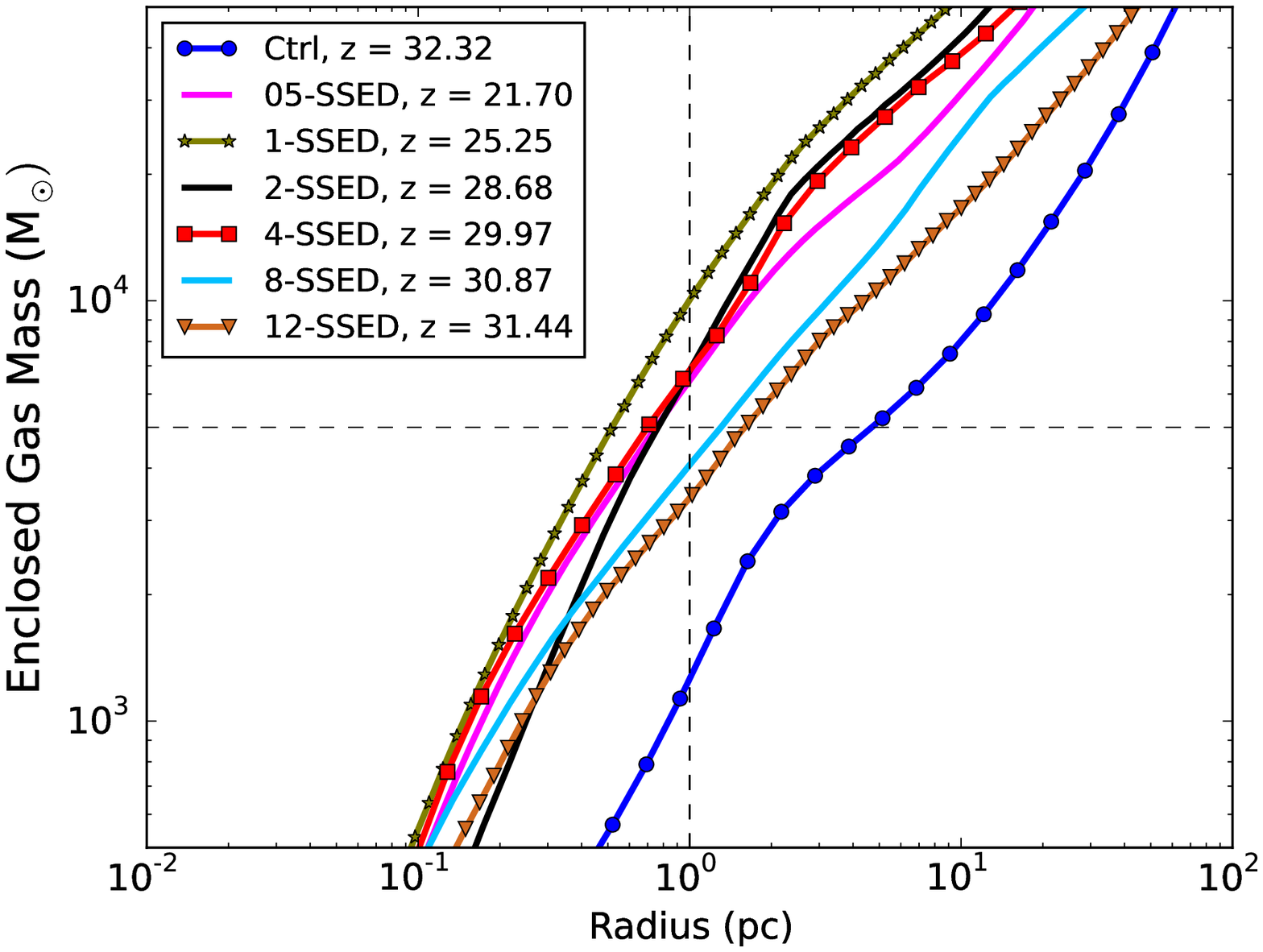}
        \includegraphics[width=9cm]{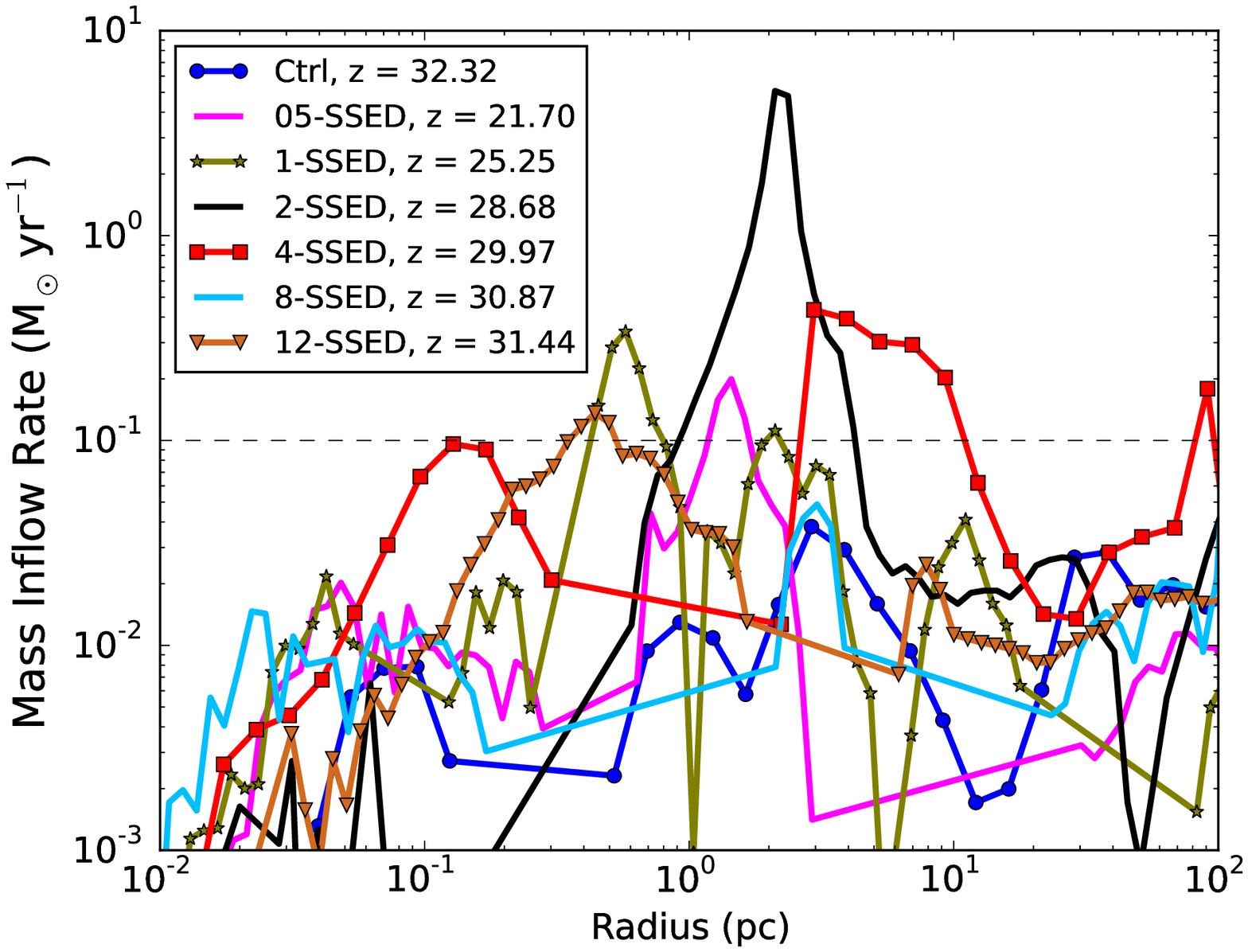}}
        \caption[]
        {\label{HaloB_EnclosedMass}
          In the left hand panel we have plotted the enclosed mass for all the 
          simulations run with a stellar SED. We have also included the 
          control run which features no radiation source for comparison. The key point is that 
          the runs with a higher radiation source have more gas at each radii. For example at a 
          radius of 1 pc there is approximately 10 times more mass at that radius in the 1-SSED compared to the 
          control case. The dashed lines are at a radius of 1 pc and an enclosed mass of 5000 \msolar and 
          are merely to guide the eye. In the right hand panel we have plotted the associated mass 
          inflow rates as a function of radius. Only the control model (Ctrl) and the 8-SSED model
          have rates which which do not exceed 0.1 \msolar yr$^{-1}$ at some radius. The 2-SSED model exceeds 
          1 \msolar yr$^{-1}$ at a radius of $\sim 2$ pc and is likely due to the state of the gas at this output.
          Overall the rates are as expected with the highest inflow rates occuring for the halos exposed to the 
          most intense radiation field. We also add a note of caution when comparing to the 05-SSED run that it is 
          not the same halo as in each of the other cases. In both panels we have ``zoomed in'' on the available ranges for 
          illustration.
        }
      \end{center} \end{minipage}
  \end{figure*}


\subsubsection{Blackbody with T = $10^5$ K}
In Figure \ref{HaloB:MultiPlot_BB1e5} we show the ray profiles for the spectrum with a blackbody 
temperature of T = 10$^5$ K. The radiation peak in this case is closer to the Lyman-Werner band 
(see Figure \ref{BlackbodySED}) and so the effect of the \molH dissociation is more clearly evident 
closer to the source - see top left panel of  Figure \ref{HaloB:MultiPlot_BB1e5} and compare to the top 
left panel of Figure \ref{HaloB:MultiPlot_BB1e4}. 
The temperature profiles are also quite different compared to the blackbody spectrum with T = 10$^4$ K. This is 
because the spectrum with a blackbody of T = 10$^5$ K both heats the gas and also disrupts its ability to 
cool by ionising hydrogen. As a result we do not observe the characteristic shock heating at the virial radius 
seen in the profile plots of Figure \ref{HaloB:MultiPlot_BB1e4} at a radius of approximately 10 - 100 pc. Instead 
the gas begins to cool due to Lyman-alpha and recombination cooling as we move away from the source. In the 
centre of the halo where the \molH fractions are able to self-shield and \molH cooling can continue unabated 
the temperature drops to T $\sim 1000$ K. These temperatures are similar to what was observed in 
Figure \ref{HaloB:MultiPlot_BB1e4} and reflects the fact that \molH shielding is able to block the effects of 
LW radiation in both cases resulting in similar levels of \molH in both cases. What is also evident from Figure 
\ref{HaloB:MultiPlot_BB1e5} is that the destruction of \molH is not a smoothly varying 
function of distance as was seen in Figure \ref{HaloB:MultiPlot_BB1e4}. In the case of the T = 10$^5$ K 
spectrum we also now have hydrogen ionising radiation. This radiation produces free electrons from the 
ionisation process. These electrons are then free to combine with the remaining neutral hydrogen or ionised 
hydrogen which is recombining. The increase in the H$^-$ fraction can increase the \molH fraction at small 
distances from the source. However, the hydrogen ionising radiation has a shorter mean free path than the 
Lyman-Werner radiation and so we see a dip again in the \molH fraction (top left panel) before it quickly 
rises again as the effects of an increasing \molH column density become more apparent.

\subsubsection{Stellar Spectrum}
In Figure \ref{HaloB:MultiPlot_SED} we show the result from the ray profiles due to the source with 
a stellar-like SED. This SED resembles a combination of the T = $10^4$ K blackbody spectrum and the 
T = $10^5$ K blackbody spectrum. The temperature profiles are quite similar to those from the T = $10^5$ K 
blackbody spectrum. However, in the case of both the source at an initial distance of 0.5 kpc and 2 kpc there 
is a clear cooling of the gas outside of the virial radius of the collapsing halo (red and green solid lines). 
The reason for this is that there is less radiation with energies greater than 13.6 eV compared to the 
blackbody spectrum with T = $10^5$ K. In the 
case of the 0.5 kpc source this is because the collapse halo is located quite far from the source - at a final 
separation of $\sim 5.0$ kpc while in the case of the source with an initial separation of 2.0 kpc the final 
separation is 2.9 kpc. In both cases the HII region surrounding the source does not reach the virial 
radius of the collapsing halo.  As a result there is a clear shock at the virial radius for these two cases 
and the gas heats up. For the case of the source with an initial separation of 1 kpc the effects of 
photo-heating and ionisation are felt up and inside the virial radius of the collapsing halo (because the final 
separation is 1.9 kpc - just inside the HII region). As with the T = $10^5$ K spectrum the \molH fraction shows 
a very strong drop close to the source before rising quickly due to the presence of free electrons which 
facilitates the production of \molHc. As the free electron fraction drops we again see a decrease in the 
\molH fraction as the catalyst (free electrons) is not available to facilitate the production of \molHc. This
effect is clearly seen in both the 05-SSED and 2-SSED models. 
The \molH fraction then increases slowly as the flux reduces before we finally enter the self-shielded region 
within approximately 0.1 pc - 1.0 pc of the central object. The values of the flux at the centre of the 
collapsing halo vary between 1 and 10 J$_{21}$ for this realistic source. The blue solid line with a flux of less 
than 1 J$_{21}$ in Figure  \ref{HaloB:MultiPlot_SED} results because the collapsing halo in this case is 
not the target halo. The collapsing halo in this case (see Figure \ref{HaloB_Projections_SED}) collapses 
at a distance of $\sim 5.0$ kpc from the source and hence the flux at the point is relatively low. 

\subsection{Increasing the source distance}
In Figure \ref{HaloB:IncreasedDistance} we show the result of ray profiling for the stellar spectrum case when
the source is moved to a distance of 4 kpc, 8 kpc and 12 kpc from the target halo respectively. The 4-SSED model 
is very similar at every scale to the 2-SSED model. However, significant differences emerge once the source 
is moved to 8 kpc and 12 kpc. In these cases the temperature is systematically lower at almost all scales
of interest. The reason for this is clear from the \molH Fraction plot in the top left panel. Between approximately 
1 pc and 50 pc from the central region the \molH fraction in each case differs by up to an order of magnitude. The increased 
level of \molH in the 12-SSED results in a lower temperature on average over that range.  The characteristics
of model 12-SSED are becoming very similar to those of the no radiation case as shown in 
Table \ref{Table:radiation_particle} reflecting the fact that at these distances the radiative flux has a rather 
small effect on the dynamics of the halo.

\subsection{Mass Inflow Rates}
In Figure \ref{HaloB_EnclosedMass} we have plotted both the mass inflow rates and the 
enclosed mass at the final output time for models with a stellar spectrum and a range of initial distances.
The mass inflow rates are calculated using the radial velocity as follows:
\begin{equation}
\dot{M}(t) = 4 \pi R^2 \rho (R) V(R)
\end{equation}
\noindent where $\dot{M}(t)$ is the mass inflow rate, R is the radius, $\rho$ is the density and 
V(R) is the radial velocity at R. For all values we choose spherically averaged quantities centred on 
the point of maximum density. While this is clearly an approximation to the true mass inflow rate, which will 
likely be episodic and possibly anisotropic, it does provide us with an insight into the 
gas dynamics as a function of radius. The anisotropic nature of our radiation source is only significant outside 
of approximately 100 pc from the centre and hence we consider here spherical profiles within the central 100 pc.
In the left hand panel of Figure \ref{HaloB_EnclosedMass}
we have plotted a radial profile of the Enclosed Gas Mass for models with initial separations of 
0.5 kpc up to 12 kpc, we have also plotted the values in the case where no radiation field is used (Ctrl model). 
The key point is that as the radiation flux increases we see a proportional increase in the enclosed gas mass 
at a given radius. If we consider the core radius to be 1 pc we see that the enclosed mass at this point is 
between approximately 10000 \msolar for the 1-SSED case down to approximately 1000 \msolar for the Ctrl case.\\
\indent Of equal importance is the mass inflow rate onto the central object which is a critical component
in determining the final outcome of the central object (see \S \ref{Sec:Discussion}). In the right hand panel 
we have plotted the mass inflow rate at the final output time for the same range of models. The inflow rates 
vary somewhat over time but generally increase as the collapse proceeds. The 1-SSED, 2-SSED and 4-SSED models 
show strong inflow rates between a few tenths of a parsec out to several tens of parsecs with an average
mass inflow rate of $\sim 0.1$ \msolar yr$^{-1}$ over that range. The values in this 
case are consistent with those of \cite{Hosokawa_2013} and \cite{Schleicher_2013} who advocate 
values of $\gtrsim 0.1$ \msolar yr$^{-1}$ are required for the formation of super-massive stars or quasi stars.\\
\indent As the flux drops (or equivalently as the distance to the source increases) the mass inflow rates 
drop systematically. It should also be noted that the value of the mass inflow rates for model 2-SSED are 
strongly peaked at a radius of $\sim 2$ pc. This is due to the dynamics of the collapse of this particular model.
A full analysis of the accretion rate onto a central object, the subsequent evolution and the associated feedback 
effects is beyond the scope of this study.

\section{Discussion} \label{Sec:Discussion}
This study has focused on examining the effect of stellar radiation from a realistic source halo on the 
collapse of a neighbouring or satellite halo. As a consequence of this study several points are worth 
noting:
\begin{itemize}
\item Our most promising candidate for forming a DCBH is the 1-SSED simulation. The halo collapses at a redshift 
of z = 25.25 with a virial temperature of $\rm{T_{vir}} \sim 9500$ K. The mass inflow rate onto the halo 
is extremely high at the time of collapse with a mass inflow rate of $\gtrsim 0.1$ \msolar yr$^{-1}$. 
The gas mass within the central core ($\sim 1$ pc) is $\sim 10000 $ \msolar and the gas is hot with temperatures
up to 10000 K at the virial radius ($\rm{R_{vir}}\ \sim 300 pc$). At the time of collapse the distance between
the central object and the radiation source is $\sim 1.9$ kpc. 

\item Using the stellar spectrum as the most realistic spectrum models 2-SSED, 4-SSED, 8-SSED are also candidates. 
However, in all of these cases the temperature and enclosed mass values drop as the distance to the 
radiation source increases. Further investigation of the collapse physics and the impact of the 
surrounding mass envelope out to distances of a few hundred parsecs will be required to clearly distinguish those 
halos which can and cannot form DCBH seeds. 

\item The ionising region can completely disrupt the collapse if the radiating halo is too close to the 
target halo. In three cases 05-T5, 1-T5 and 05-SSED we saw that the
target halo is unable to collapse and a different minihalo approximately 5.0 kpc from the radiating 
source ultimately collapses first. The fact that the target halo is enveloped within the HII region of the 
source halo means that its collapse at a later stage is unlikely (as long as the HII region remains). Furthermore, 
due to the close proximity of the target halo to the collapse the target halo is likely to become polluted 
by metals due to winds from the radiating source particularly for the cases where the halo is placed at the 
shorter distances of 0.5 kpc.  
 
\item For the cases where the target halo is completely disrupted a halo on the edge of the HII region 
collapses which is different from the original target halo. The virial temperature in each case is greater 
than 8000 K and therefore atomic cooling is operational. Taking the 05-SSED as the best example, the 
characteristics of the 05-SSED case are somewhat similar to the 1-SSED case. The major difference is that the 
flux is lower by a factor of 10 (because the collapsed halo is now $\sim 5.0$ kpc from the source) and the 
temperature of the gas at a radius of approximately 200 pc is also significantly higher. 
This impacts on the effects of the LW radiation at this scale, the  \molH fraction between $\sim 10$ pc and 500 pc
is in fact higher in the 1-SSED case contrary to what might be naively expected. This is a direct result of the 
higher gas temperature of the gas at this scale. 

\item The mass inflow rates found in our simulations have peaks that are greater than 0.1 \msolar  yr$^{-1}$. 
In particular models 1-SSED, 2-SSED and 4-SSED have sustained average inflow rates of $\sim  0.1$ \msolar  yr$^{-1}$
over a decade or more in radius. These values for the mass inflow rate compare well to what has 
been suggested is required for super-massive star or quasi-star formation by \cite{Hosokawa_2013} and
\cite{Schleicher_2013}. However, we do not attempt, in these simulations, 
to follow the collapse to very small scales and lack the necessary resolution and detailed physics to 
make firm predictions on the characteristics of the final object. Recent work by \cite{Hosokawa_2015}
also indicates that short accretion bursts temporarily exceeding  0.01  \msolar  yr$^{-1}$ may be all that is 
required to induce massive primordial star formation which could in turn result in DCBH seeds. It would appear
from our simulations that achieving this criteria should at least be feasible.  
 
\item The flux at this redshift is limited by the available growth time.  Our stellar flux was limited 
to $1.2 \times 10^{52}$ photons per second based on a stellar mass of $10^5$ \msolar at z = 20. Achieving 
significantly higher fluxes at this redshift is unlikely and the values of $J$ found in our simulations 
represent likely values at this epoch ($J \sim 10\ \rm{J}_{21}$ for the 1-SSED model). Moreover, in our simulations 
the source radiates at this flux starting from $z = 40$ meaning that this is a rather optimistic and somewhat 
idealised case. More realistic cases (even rare ones) would start off with much lower fluxes. Our values are 
not high enough to halt \molH cooling completely especially towards the centre of the 
halo and as a result the temperature of the gas found in the centres of halos are well below those found in 
atomic cooling haloes by approximately an order of magnitude. Hence, the formation of a truly atomic core, 
if required for DCBH seed formation \citep[see][]{Latif_2015b}, may need to wait until lower redshifts where 
a combination of a background and nearby source can generate a high enough $J$.

\item The recent identification of a very bright \lya emitter discovered at z $\sim 6.6$ \citep{Matthee_2015,
Sobral_2015} has been followed up by some theoretical studies proposing that the \lya source may be a DCBH
\citep{Pallottini_2015, Agarwal_2015, Hartwig_2015b}. In particular \cite{Agarwal_2015} find that in modelling 
CR7 that DCBH formation occurs in the range $19 < z < 23$. In our simulations by turning on our initial 
(atomic halo) source at $z = 40$ we find that we can form near-atomic cooling haloes between z $\sim 20$ and 
z $\sim 26$. These haloes have virial temperatures between 7000 K and 10000 K. As discussed above model 1-SSED 
would provide a good fit to the results of \cite{Agarwal_2015}. However, the core is primarily cooled by 
\molH and the fate of the central object is unclear. Further study of the detailed physics of the collapse
in this scenario is the next logical step. 

\item To simulate a nearby atomic cooling halo hosting a galaxy we use a radiation particle to 
model the emission. We do this so as to increase the flexibility of our parameter study and this allows us to 
easily alter the distance from the radiating galaxy to the target halo. It should be noted however that we 
miss some important considerations in this case. In particular these systems, given their close proximity, are 
likely bound systems and this is not taken into account in our model. Nor is the fact that as the 
evolution proceeds these systems may decrease in separation and may in fact merge. This scenario will be 
be investigated in an upcoming study (Regan et. al in prep). 

\end{itemize}

\section{Conclusions} \label{Sec:Conclusions}
Using a multi-frequency ray-tracing scheme we have investigated the impact of three different spectral shapes
on a collapsing halo. We have used a blackbody spectrum with T = $10^{4}$ K, a blackbody spectrum with
T = $10^{5}$ K and a spectrum with a spectral shape consistent with that produced by a 
galaxy at z $\sim 20$. In each case the photon emission rate is unchanged and has been set 
to $1 \times 10^{52}$ photons per second consistent with simulations of the first galaxies \citep{Chen_2014}. 
To model the emitting galaxy which is supposed to reside close to the collapsing 
halo we have developed a radiation particle which we place at fixed initial distances from the halo. We vary these
initial distances between 0.5 kpc and 12 kpc from the halo to investigate the impact of the radiation on the 
collapsing halo. The radiation covers the range 0.1 eV up to 60 eV. \\
\indent Using the stellar spectrum as our fiducial result we find that placing the source too close to the 
collapsing halo (distance less than 1.0 kpc) results in the total disruption of the collapsing halo due to 
photo-ionisation of hydrogen. As the source is moved further from the halo we find that at distances greater than 
4 kpc the intensity of the radiation impacting the halo is at or below what would be expected from the background
at high redshift. The resulting virial temperatures of the collapsed halo are well below that where atomic 
processes dominate and as a result the halo is unlikely to be a candidate for forming direct collapse seeds. \\
\indent At initial separations of 1 kpc (and 2 kpc) we are able to form halos with  virial 
temperatures greater than 6000 (4000) K. Crucially, in the case of the source at a separation of 1 kpc (model
1-SSED) the object that forms at the centre of this halo is surrounded by a large envelope of hot gas with an 
enclosed mass nearly 10 times greater than what is observed in the halos subjected to weaker radiation fields. 
Furthermore, the mass inflow rates observed for this halo have average values greater than $\sim 0.1$ 
\msolar yr$^{-1}$ out to several tens of parsecs. These environmental conditions favour the formation of 
extremely massive primordial stars \citep{Hosokawa_2015, Hosokawa_2013, Schleicher_2013}, in neighbouring 
satellite halos, and could potentially be the ideal environment in which to form massive black hole seeds. 
Further investigation of the detailed physics of the collapse, given these environmental conditions, 
is now required to determine the exact nature of what object(s) form and whether they can then 
collapse to form a massive black hole seed. 

\section*{Acknowledgements}
\noindent J.A.R. and P.H.J. acknowledge the support of the Magnus Ehrnrooth Foundation, the Research
Funds of the University of Helsinki and the Academy of Finland grant 1274931.
This work was also supported by the Science and Technology Facilities Council [ST/F001166/1].
J.H.W. acknowledges support by NSF and NASA grants AST-1333360 and
HST-AR-13895.001.  The numerical simulations were performed on facilities 
hosted by the CSC -IT Center for Science in Espoo, Finland, which are financed by the 
Finnish ministry of education. Computations described in this work were performed using the 
publicly-available \enzo code (http://enzo-project.org), which is the product of a collaborative 
effort of many independent scientists from numerous institutions around the world.  Their 
commitment to open science has helped make this work possible. The freely available astrophysical 
analysis code YT \citep{YT} was used to construct numerous plots within this paper. The authors 
would like to express their gratitude to Matt Turk et al. for an excellent software package.
All modified source code used in this work is freely available from J.A.R.'s bitbucket account and 
is in the process of being pushed to the Grackle and Enzo mainlines. Email J.A.R for more information 
on acquiring the updated source code if required.  Finally, the authors would like to thanks an 
anonymous referee for a detailed and helpful report.

\end{document}